\documentclass[aps,prb,twocolumn,superscriptaddress,
               amsmath,a4paper,floatfix]{revtex4}

\usepackage{graphicx}
\usepackage{dcolumn}
\usepackage{bm}
\usepackage{dcolumn}
\usepackage{subfigure}
\usepackage{color}
\usepackage{amsmath}
\usepackage{amssymb}
\usepackage{ulem}
\usepackage{psfrag} 

\graphicspath{{../}}
\begin{document}
\title{Perturbative Quantum
Monte Carlo Study of LiHoF$_4$ in a Transverse Magnetic Field }

\author{S.M.A Tabei} 
\affiliation{Department of Physics and Astronomy, 
University of Waterloo, Waterloo, Ontario, N2L 3G1, Canada}

\author{M.J.P. Gingras}
\affiliation{Canadian Institute for Advanced
 Research, 180 Dundas St. W., Toronto, Ontario, M5G 1Z8, Canada}

\affiliation{Department of Physics and Astronomy,
University of Waterloo, Waterloo, Ontario, N2L 3G1, Canada}

 \author{Y.-J. Kao}
 \affiliation{Department of Physics  and Center for
Theoretical Sciences, National Taiwan University, Taipei 10617, Taiwan}
\author{T. Yavors'kii}

\affiliation{Department of Physics and Astronomy,
University of Waterloo, Waterloo, Ontario, N2L 3G1, Canada}

\begin{abstract}
 Results from a recent quantum Monte Carlo (QMC)
study 
(P.B. Chakraborty {\it et al.}, Phys. Rev. B {\bf 70}, 144411 (2004))
study of the LiHoF$_4$ Ising magnetic material in an external 
transverse magnetic field 
 $B_x$ show a discrepancy with the experimental results, even for small $B_x$ where quantum fluctuations are small.
This discrepancy persists
asymptotically close to the classical ferromagnet to paramagnet phase transition. In this paper, 
 we numerically reinvestigate the temperature $T$, versus transverse field phase diagram of LiHoF$_4$ in 
 the regime of weak $B_x$. In this regime, starting from an effective low-energy spin-$1/2$ description 
 of LiHoF$_4$, we apply a 
cumulant expansion to derive an effective temperature-dependent classical
 Hamiltonian that incorporates perturbatively the small quantum fluctuations in the vicinity of the classical 
 phase transition at $B_x=0$. Via this effective classical Hamiltonian, we study the $B_x-T$ phase diagram
 via classical Monte Carlo simulations. In particular, we investigate the influence on the phase 
 diagram of various effects that may be at the source of the discrepancy between the previous QMC results 
 and the
 experimental ones. For example, we consider two different ways of handling the long-range dipole-dipole 
 interactions and explore how  the $B_x-T$ phase diagram is modified when using different microscopic 
 crystal field Hamiltonians.
 The main conclusion of our work is that we fully reproduce the previous QMC results at small $B_x$. Unfortunately,
 none of the modifications to the microscopic Hamiltonian that we explore are able to provide a $B_x-T$ phase diagram
 compatible with the experiments in the small semi-classical $B_x$ regime.
\end{abstract} 
\maketitle
\section{Introduction}
\subsection{Transverse Field Ising Model}
Phase transitions from order to disorder are most commonly driven by thermal fluctuations.
 However,
near absolute zero temperature, a system can, via {\it quantum} fluctuations associated 
with the Heisenberg uncertainty principle, undergo a quantum phase transition (QPT) 
\cite{Sachdev,Sondhi}.
The transverse field Ising model (TFIM) is perhaps the simplest model that exhibits a
QPT~\cite{Sachdev,Pfeuty,Sen}.
 This model was first proposed by de Gennes
to describe proton tunneling in ferroelectric systems~\cite{deGennes}. 
The Hamiltonian of the TFIM is given by 
\begin{eqnarray}
H_{\rm TFIM}=-\frac{1}{2}\sum_{i,j}J_{ij}\sigma_i^z\sigma_j^z 
 -\Gamma\sum_i\sigma_i^x~,
 \label{TFIM}
\end{eqnarray} 
where $\sigma_i^\mu$ ($\mu=x,y,z$) are the Pauli matrices. 
Since $\sigma_i^x$ and $\sigma_i^z$ do not commute, a nonzero field $\Gamma$, transverse to the Ising $\hat z$ direction,
causes quantum tunneling between the spin-up and spin-down eigenstates of
$\sigma_i^z$, hence causing quantum spin fluctuations. These fluctuations decrease the critical temperature $T_c$ 
at which the spins develop long-range order. In the simplest scenario, where $J_{ij}>0$, 
the ordered phase is ferromagnetic~\cite{Pfeuty,Sen}.  
At a critical field $\Gamma_c$, $T_c$ 
 vanishes, and a quantum phase transition between the quantum paramagnet (PM) 
and a long-range ordered ferromagnetic state occurs. The $H_{\rm TFIM}$ can be generalized by considering 
$J_{ij}$ as quenched (frozen) random interactions. Competing ferromagnetic $J_{ij}>0$ and 
antiferromagnetic $J_{ij}<0$ couplings generates random frustration. For a three dimensional
 case, the system freezes into an (Ising) spin glass state at a spin glass
 critical temperature $T_g$~\cite{Binder,Ballestros}.
 Similarly to the previous example, $T_{g}(\Gamma)$ decreases as $\Gamma$ 
 is increased until,
at $\Gamma=\Gamma_{c}$, a quantum phase transition between 
a quantum paramagnet and a spin glass phase occurs.
 Extensive numerical
studies have found the QPT between a quantum
paramagnet and a spin glass phase~\cite{Rieger-PRL,Rieger-PRB,Guo} 
to be quite interesting due to the
occurrence of Griffiths-McCoy singularities~\cite{Griffiths,McCoy}.

 \subsection{LiHo$_x$Y$_{1-x}$F$_4$}

 The magnetic insulator LiHoF$_4$, with a magnetic field $B_x$ applied
perpendicular to the Ising $z$ direction of the Ho$^{3+}$ magnetic moments, 
is a well known example of a physical 
realization of the transverse field Ising model~\cite{Bitko, Wu, Ronnow,Hansen}. 
In  LiHoF$_4$ the predominant
 $J_{ij}$ interaction between the Ho$^{3+}$ ions
 is the long range interaction between magnetic dipoles which
 decays as $1/r_{ij}^3$,
 where $r_{ij}$
 is the distance between the $i$ and $j$ ions. The sign of $J_{ij}$ depends
   on the position of $j$ respect to $i$.
 The existence of a large crystal field
 anisotropy on the magnetic Ho$^{3+}$ ions~\cite{Hansen} causes the system to behave as a classical 
 Ising system with dipolar interactions for zero applied magnetic field $B_x$.
 The reason is that the single ion crystal field ground state is an Ising doublet, 
 meaning that the matrix elements of the raising and lowering angular momentum operator 
 ${\rm J}^{\pm}$ vanish within the space spanned by the two states of the doublet. 
  The Ising 
 direction is parallel to the \textbf{c} axis of the body centered tetragonal structure of
  LiHoF$_4$.
  In zero applied magnetic field $B_x$, the system is well described by a low-energy effective spin-$1/2$ 
  classical dipolar 
  Ising model~\cite{Eastham,Tabei-Vernay}. Because the energy gap between the ground doublet and the first
   excited singlet is fairly large compared to the $J_{ij}$ couplings, there is  little quantum
    mechanical admixing between
  the ground doublet and the excited state induced by the interactions~\cite{Eastham}. However, a nonzero $B_x$ 
  admixes the ground doublet with the excited singlet and splits the ground doublet.  
  It is this energy splitting which corresponds to the effective transverse field $\Gamma$ in the TFIM
  description of LiHoF$_4$ in nonzero $B_x$~\cite{Bitko,Tabei-Vernay,prabuddha}.
  
   The Ho$^{3+}$ ions may be substituted~(i.e.~randomly diluted)
by
 non-magnetic yttrium (Y$^{3+}$) ions, with very little lattice distortion.
This allows one to study the effects of disorder on LiHo$_x$Y$_{1-x}$F$_4$ as an example of 
a diluted Ising model. 
Depending on the concentration $x$ of magnetic ions, the low temperature phase 
is either ferromagnetic~\cite{Bitko,Silevitch} or spin glass~\cite{Reich,No-SG,Jonsson}. 
Interestingly,
 paradoxical behaviors are observed when a
  transverse magnetic field is applied to LiHo$_x$Y$_{1-x}$F$_4$, with $x<1$. 
In the ferromagnetic regime, ($0.25 < x <1.0$), 
when $B_x=0$, a mean-field behavior $T_c(x) \propto x$ for 
the paramagnet to ferromagnet temperature transition is observed.
However, in nonzero $B_x$, with increasing $B_{x}$, $T_c(B_{x})$ decreases faster than 
mean field theory predicts~\cite{Brooke-thesis}.
For $B_x=0$, when LiHo$_x$Y$_{1-x}$F$_4$ is diluted below $x\approx0.25$, 
  a conventional spin glass transition is observed~\cite{Wu,No-SG,Jonsson}. The signature of the spin glass transition
  is the divergence of the nonlinear magnetic susceptibility
$\chi_3$
  at $T_g$~\cite{Mydosh}.   
However, surprisingly, $\chi_3(T)$ becomes less singular as $B_{x}$ is increased from ${B_{x}=0}$,
suggesting that no quantum phase transition between a PM and a SG
state exists as $T\rightarrow 0$ \cite{Wu,Wu-thesis}.
Recently, theoretical studies~\cite{Schechter-PRL2,Tabei-PRL,Schechter-JPC,Tabei-Vernay}
have suggested that for dipole-coupled Ho$^{3+}$ in diluted LiHo$_x$Y$_{1-x}$F$_4$,
nonzero $B_{x}$ generates  longitudinal 
(along the Ising $\hat z$ direction) random fields 
that couple to the magnetic moment and (i) lead 
to a faster decrease of $T_c(B_x)$ in the ferromagnetic 
regime and (ii) destroy the 
paramagnet to spin glass transition in LiHo$_x$Y$_{1-x}$F$_4$ samples that otherwise
show a SG transition when $B_x=0$~\cite{No-SG,Jonsson}. 
Recently, for the ferromagnetic regime, the influence of
 these induced random fields  on the 
 behavior of the linear magnetic susceptibility $\chi$ in 
 the presence of an external transverse magnetic field
  has been experimentally studied~\cite{Silevitch}.
When LiHo$_x$Y$_{1-x}$F$_4$ is highly diluted (e.g. LiHo$_{0.045}$Y$_{0.955}$F$_4$), very interesting 
and peculiar behaviors are observed. AC susceptibility data show
that the distribution of relaxation times {\it narrows} upon cooling below
300 mK ~\cite{Reich,Ghosh-Science,Ghosh-Nature}.
This behavior is quite different from that observed in conventional
spin glasses, where the distribution of relaxation times broadens upon
approaching a spin glass transition at $T_{g}>0$ ~\cite{Binder,Mydosh}. 
This so-called {\it antiglass} behavior has been interpreted as evidence that
the spin glass transition in  LiHo$_{x}$Y$_{1-x}$F$_4$ disappears
at some nonzero $x_c>0$ \cite{No-SG,Jonsson}. 
This is in contrast with theoretical arguments~\cite{Stephen}
which argue that, because of the long-ranged $1/r^3$ nature of dipolar
interactions,  classical dipolar Ising spin glasses should have 
$T_g(x)>0$ for all $x>0$.
However, recent numerical~\cite{Snyder,Biltmo} 
and experimental
works~\cite{Jonsson} claim that a finite temperature paramagnetic to spin glass phase transition
 may not occur for $x$ as large as $x_c\approx$~0.2.
\begin{figure}[tbp]
\includegraphics[width=\columnwidth,angle=0]{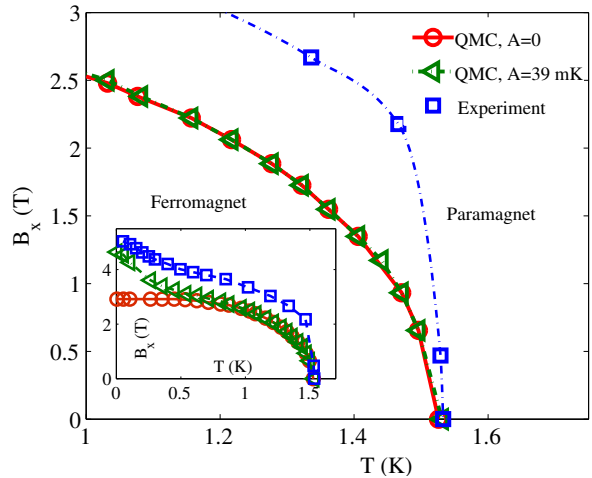}
\caption{\small{ The discrepancy between the experimental~\cite{Bitko} phase diagram 
of LiHoF$_4$
and quantum Monte Carlo (QMC) simulations using stochastic series expansion for small $B_x$ from Ref.~[\onlinecite{prabuddha}]. The whole 
phase diagram is shown in the inset. At low temperature and
high $B_x$, neglecting the large hyperfine interaction $A$, generates  
a significant discrepancy between 
the experimental 
quantum critical point and the one obtained from simulation. However, 
 at low $B_x$ and close to the classical critical point,
the hyperfine interaction is not a quantitatively important parameter. Other possibilities
for the origin of this discrepancy have to be invoked in this regime. 
\label{whole}}}
\end{figure}

\subsection{LiHoF$_4$ as a TFIM}\label{int-LiHo}

In addition to the phenomena arising in the diluted regime
of LiHo$_x$Y$_{1-x}$F$_4$, 
the $x=1$ regime also turns
out to be interesting. 
There still exist problems for the pure LiHoF$_4$, 
requiring the properties of 
this system in nonzero $B_x$ to be re-investigated more thoroughly.
Perhaps surprisingly, it
 is just recently that the   
  properties of LiHoF$_4$ in a transverse 
 external magnetic field have been studied 
 in quantitative detail 
starting from a truly microscopic spin Hamiltonian~\cite{prabuddha}. 
 In Ref.~[\onlinecite{prabuddha}], which reported results from a quantum Monte Carlo (QMC) study
  using the stochastic series expansion (SSE) technique~\cite{sandvik},
 a general qualitative agreement between the microscopic model and 
 experimental data~\cite{Bitko} was obtained. However, as illustrated in Fig.~\ref{whole}, 
 there is significant quantitative discrepancy between
 the Monte Carlo results of Ref.~[\onlinecite{prabuddha}] and the experimental data of Ref~[\onlinecite{Bitko}]. 
In particular, the
discrepancy between experiment and QMC results persists asymptotically close to the
   classical ferromagnetic to paramagnetic phase transition, where 
 $B_x/T_c$ and quantum fluctuations are {\it perturbatively small}. 
For very low temperatures
 and high $B_x$, it is crucial to consider the hyperfine interaction in order to explain the behavior of the
 phase diagram close to the quantum critical point~\cite{Bitko,prabuddha,Schechter}. 
However, for very small $B_x/T_c$, 
the numerical results shown in Fig.~\ref{whole} indicate that the 
effect of the 
 hyperfine interaction is not important close to the classical transition at $T_c$.
   
 It was suggested in Ref.~[\onlinecite{prabuddha}] that this discrepancy between 
  simulation and experiment, close to the classical transition, may be related to
  some uncertainty 
 in the crystal field parameters (CFP) used in the crystal field Hamiltonian, which  enters in
 the TFIM description of LiHoF$_4$, and which is simulated via QMC.
  Indeed a number of 
 CFP sets obtained from different experimental works, such as susceptibility measurements~\cite{Hansen},
 neutron scattering~\cite{Ronnow}, and electron paramagnetic resonance  
 experiments~\cite{Malkin}, provide rather different values for the CFP. Specifically,
  different CFP would lead to different field ($B_x$) dependent effective coupling parameters
   in the TFIM description of LiHoF$_4$, which would result in different $B_x$ vs $T_c$ phase diagrams.
   
   Yet, there are other factors of strictly computational nature which may be at the origin of the 
discrepancy illustrated in Fig.~\ref{whole}. 
 For example, because of the difficulties associated with dipolar interactions, calculations incorporating
  long-range dipolar interactions need to be performed quite carefully.
  Because of the long-range nature and 
angular dependence of dipolar interactions, the dipolar sum 
$U(i)=-1/N\sum_j(1-3\cos^2\theta_{ij})/r_{ij}^3$ is conditionally convergent~\cite{Luttinger,conditionally,comment}, i.e 
the value of the sum depends on the shape of the external boundary of the system studied.
Here, $r_{ij}$ is the distance between site $i$ and $j$, and $\theta_{ij}$ is the angle between 
$r_{ij}$ and the Ising spin axis.

The  conditional convergence of dipolar sums has been studied by
 Luttinger and Tisza~\cite{Luttinger}. They performed the dipolar sum for 
a number of spin structures 
for systems with different external boundary shapes. 
For example, they considered an infinitely large system of dipoles on a
body centered cubic lattice. They found that when the external boundary is spherical, the ground
state is antiferromagnetic, while it is ferromagnetic for a needle-shaped sample. 
Later, Griffiths rigorously proved
  that  
 for  zero external field
  the free energy for a dipolar lattice system has to be independent of the sample shape
  in the thermodynamic limit~\cite{Griffiths2}.  The immediate consequence of Griffiths' theorem is that  
  in zero external field, the net magnetization of the sample 
  has to be zero. Otherwise, the field caused
by the magnetic moments sitting on 
   the boundary of the sample would  
    couple to the dipolar moments of the sample, making the free energy shape dependent.
   Therefore, as a result of Griffiths' theorem~\cite{Griffiths2}, domains must form
in the sample, 
   such
   the total magnetization of the sample is zero in the thermodynamic limit.
  Griffiths' theorem is at variance with Luttinger and Tisza~\cite{Luttinger} results
  because, in their work,
the spin configurations were assumed uniform, and domain formation was neglected.
This discussion emphasizes the complication of studying
systems with dipolar interactions and the caution which should be taken 
while dealing with such systems 
(e.g. the choice of the boundary geometry, 
boundary conditions and
and the shape of the domains.)
 Finite size effects is another issue that needs to be handled quite carefully 
  in  systems where ions interact via long-range interactions.        
 
 There are different ways to incorporate dipolar interactions in a computationally efficient way.
  The method implemented 
 in Ref.~[\onlinecite{prabuddha}] is the reaction field method~\cite{RF}, 
 which truncates the sum of the long-range 
 interactions at the boundary of a sphere. The dipoles outside the sphere
  are treated in a mean-field fashion.
 Due to the semi mean-field nature of this method, the reaction field method 
 overestimates
the critical temperature.
 In the presence of quantum fluctuations, this overestimation is still at play and 
 can possibly influence
 the $B_x$-$T_c$ phase diagram as well. 
 The Ewald summation method~\cite{Zeeman,huang,Leeuw, spinice,Melko} is 
another method to treat the long-range dipolar interactions.
 In the Ewald 
 summation method, a specified volume is periodically replicated. Then, by 
 summing two convergent series 
effectively representing the dipolar
interactions between magnetic moments
$i$ and $j$, and all the periodically 
   repeated images of $j$, an effective dipole-dipole interaction
  between two arbitrary magnetic moments 
 $i$ and $j$ within the finite size sample to be numerically simulated is derived. 
 From a general perspective, it would appear quite
   worthwhile to investigate the applicability and usefulness of the Ewald summation method to determine 
   the low $B_x$ vs $T_c$ phase diagram of LiHoF$_4$. Indeed, the Ewald summation method, unlike the reaction 
   field one,
   is less prone to mean field
   over-estimations, and can be used as another methodology to probe the 
   LiHoF$_4$ problem via simulations~\cite{Biltmo}.      

 Another factor whose influence 
on the $B_x-T$ phase diagram that should be studied is the  
  nearest neighbor exchange interaction $J_{\rm ex}$ in LiHoF$_4$.
 The strength of  $J_{\rm ex}$, 
 which is
expected to be comparable to the dipolar interactions for a $4f$ ion such as Ho$^{3+}$, is unknown. 
 The strength can be determined such that the classical critical temperature
  matches the experimental value for $B_x=0$.
 The estimated value of $J_{\rm ex}$ is highly sensitive to the method 
  used to handle the external boundaries and finite size effects in simulations, 
both of which have significant effects
when using the 
reaction field (RF) 
method, as already found
  in Ref.~[\onlinecite{prabuddha}]. 

 \subsection{Scope of the Paper}

 The above discussion should make it clear that there are two 
rather distinct
avenues to
 pursue in order to seek an explanation for
 the discrepancy between the experimental~\cite{Bitko} $B_x$ vs $T_c$ phase diagram of LiHoF$_4$ and 
 the one obtained via QMC~\cite{prabuddha}. One avenue, is that the current microscopic model 
 is incomplete. As mentioned above, and suggested in Ref.~[\onlinecite{prabuddha}], one possible 
 source for this incompleteness may be an inaccurate set of CFP. Another possible source
 is that other interactions other than long-range magnetic dipolar interactions 
 and nearest neighbor exchange may be at play~\cite{JensenBook}. 
 Examples of other interactions include higher order multipole interactions and 
 virtual phonon exchange~\cite{JensenBook}.
 The other avenue is related to the 
ensemble of computational pitfalls and insuing numerical errors that may 
arise
 when one deals with long range dipolar interactions
 through simulations.
Therefore,
 before one delves into 
 exploring a more complex microscopic Hamiltonian,
 there is a clear need to re-investigate the ``simpler" problem 
that solely considers
long-range dipole-dipole 
 interactions and nearest neighbor exchange.

 
 In this work we aim to scrutinize the individual role of each 
 of the computational issues as potential culprits for the discrepancy observed in 
 Fig.~\ref{whole}.
 Because QMC and experiment do not match at $B_x/T_c\rightarrow0$, we have developed
 a tool that allow us to achieve the goal in an efficient and computationally simple way.    
  Since this discrepancy appears at low enough $B_x$
 near the classical $T_c$, where quantum fluctuations are {\it perturbatively}
  small, we can expand 
 the partition function $Z$ in terms of the transverse magnetic field $B_x$,
  and recast the partition function as a sum over strictly classical states, using 
  a new effective,
albeit temperature dependent, {\it classical} Hamiltonian $H_{\rm eff}(T)$.
 In $H_{\rm eff}(T)$, the quantum effects are incorporated
   perturbatively, giving us the ability to calculate all thermodynamical quantities in presence of
   small quantum fluctuations within a classical Monte Carlo method. Therefore 
   classical Monte Carlo simulations can be easily performed using $H_{\rm eff}(T)$ 
   in a very simple way, without the need to perform complicated 
QMC \cite{sandvik, prabuddha} simulations when interested in a regime with 
   weak quantum fluctuations~\cite{SSE}. Therefore, we can focus on the region close to
    the classical transition 
   and investigate the different possible origins of the discrepancy in detail.
   
   In summary, (i) the 
complexity of the QMC SSE method, (ii) the problematic conditional convergence
 of dipolar lattice sums, (iii) the question of controlled finite size effects and its role on the consistent determination of the nearest-neighbor exchange $J_{\rm ex}$, and (iv) the possible sensitivity of the $T_c(B_x)$
 dependence on the choice of the CFP altogether warrant
 a new numerical investigation of the $T_c(B_x)$ phase diagram in
 the LiHoF$_4$ transverse field Ising material.
Below, we will show that
  either fortunately or unfortunately, depending on one's disposition, the factors proposed
 in Section \ref{int-LiHo} as the possible origins
 of the discrepancy between experiment and simulation (see Fig.~\ref{whole})
are apparently not the issue. Therefore, the origin of the discrepancy 
 remains unexplained. However, the perturbative cumulant Monte Carlo
tool that we have devised can be used effectively to search for the cause of discrepancy.
 Without it, the discovery of the irrelevance of the above factors through a classical
Monte Carlo simulation would have been a more CPU time consuming burden.
 Ultimately, the same tool can also be used to explore the role of the small $B_x$ when
  $x\neq 0$~\cite{Silevitch, Schechter-PRL2, Tabei-PRL, Schechter-JPC}. 
Indeed, 
  constructing the whole $x$-$T_c(B_x)$ phase diagram in the ``small $B_x$" 
vicinity of the classical $x$-$T_c$ phase diagram by performing solely 
classical Monte Carlo was an original key       
motivation for the development of the method presented in this paper.

 The rest of the paper is organized as follows. In Sec. II, we review the crystal structure
  and the physical properties of LiHoF$_4$ in a  transverse field $B_x$ and the effect 
  of the choice of crystal field potential 
  on the magnetic low energy states. In Sec. III, we introduce the full microscopic Hamiltonian of LiHoF$_4$.
We  
  discuss how, 
  for low energies, an effective spin-$1/2$ Hamiltonian for LiHoF$_4$ can be constructed,
   and explain how one 
 can picture LiHoF$_4$ in nonzero $B_x$ as a dipolar TFIM. 
   We then discuss how a semiclassical effective Hamiltonian is derived from the TFIM
   Hamiltonian by incorporating the transverse field term perturbatively via a cumulant expansion.
   In Sec. IV, we employ 
 the semiclassical effective Hamiltonian obtained in the previous 
   section in  classical  Monte Carlo simulations for small $B_x$. 
   We discuss the results obtained using either the reaction field 
 or Ewald summation method for the 
    long-range dipole interactions. We discuss how $J_{\rm ex}$ is estimated and investigate
    the sensitivity of the determined value  upon 
 the choice of  the numerical method.
    Finally, we compare the $B_x$-$T_c$ phase diagrams originating from two
     different sets of  crystal field parameters. Section V  
     summarizes our results.     
The paper also contains three appendices. Appendix A discusses details pertaining
to the crystal field Hamiltonian. Appendix B gives some of the intermediate
steps needed to construct the effective classical Hamiltonian
${\cal H}_{\rm eff}(T)$. Finally, Appendix C give the formulae
needed to calculate physical thermodynamic quantities when doing classical
Monte Carlo simulations with ${\cal H}_{\rm eff}(T)$.

\section{Structure and Crystal Field}\label{Sec-2}

 The magnetic material LiHoF$_4$ undergoes a second-order 
 phase transition from a paramagnetic to a ferromagnetic state
  at a critical temperature of 1.53~K~\cite{Bitko,Hansen}. The critical 
 temperature can be reduced by applying a magnetic field $B_x$ transverse
to the Ising easy-axis direction. The magnetic field induces
quantum fluctuations such that beyond a critical field of $B_{x}^c\approx 4.9$~Tesla, 
the system displays a quantum phase transition from a ferromagnetic state to a 
quantum paramagnetic state at zero temperature~\cite{Bitko}.
The magnetic properties of LiHoF$_4$ are due to Ho$^{3+}$ rare earth magnetic ions.
The electronic ground state of Ho$^{3+}$ is $4f^{10}$, which gives small exchange 
coupling~\cite{prabuddha,Mennenga,Beauv}, 
such that the predominant
magnetic interaction between the Ho$^{3+}$ ions are long-range magnetic dipole-dipole
 interactions.    
 Hund's rules dictate that the total angular momentum of a free ion Ho$^{3+}$,
 ${\rm J} = 8$ (${\rm L} = 6$ and  ${\rm S} = 2$) 
and the electronic ground state configuration is $^5I_8$. 
 LiHoF$_4$ is a compound with space-group 
$C^{6}_{4h}\left(I4_{1}/a \right)$ and lattice parameters 
$a=b=5.175\AA$, $c=10.75\AA$, and has
4 Ho$^{3+}$ ions per unit cell positioned at 
$(0,0,1/2)$, $(0,1/2,3/4)$, $(1/2,1/2,0)$ and $(1/2,0,1/4)$~~\cite{Mennenga}. 
 The crystal has $S_4$ symmetry,
which means the lattice is invariant with respect to a $\frac{\pi}{2}$ rotation about the $z$ axis 
and reflection with respect to the $x-y$ plane.\\

In the crystal structure, the Ho$^{3+}$ ions are surrounded by F$^-$ ions, which create
a strong crystal electric field with $S_4$ symmetry.
This crystal field lifts the 17-fold degeneracy of the $^5I_8$ configuration giving a non-Kramers
ground state doublet. The next excited state is a singlet with an energy gap of $\approx 11$~K 
above
the ground state doublet~\cite{Ronnow,Hansen,Malkin,11K}.
 The crystal field Hamiltonian and the crystal field parametrization is discussed 
 in more detail in Appendix \ref{A}. 
 Holmium is an isotopically
pure element with nuclear spin ${ \rm I} = 7/2$, which is coupled to the electronic
spin ${\mathbf J}$ via the hyperfine contact interaction $A\mathbf{I}\cdot\mathbf{J}$,
 where $A\approx39$~mK~\cite{Mennenga,Magarino}.
\begin{figure}[htp]
\center
\includegraphics[width=5 cm,angle=0,keepaspectratio]{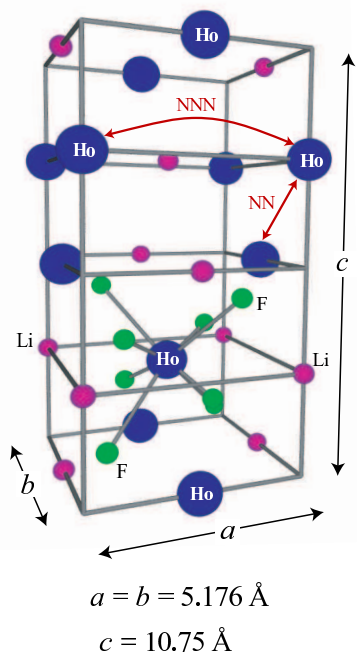}
\caption{The crystal structure of LiHoF$_4$. \rm NN identifies the first nearest neighbors
and N\rm NN identifies the next nearest neighbors}
\label{fig:crystal}
\end{figure}
%
\section{ Effective Theory of ${ \rm \bf LiHoF_4}$ for the Low ${\bf B_x/T_c}$ Regime }

In this section we derive an effective model suitable for describing LiHoF$_4$
in a small transverse magnetic field regime, where $B_x/T_c\rightarrow0$ ($T_c$ is the critical 
temperature when $B_x=0$). The simplicity gained using an effective theory
gives us the ability to capture the essential
physics, and to easily reinvestigate the influence of the 
different parameters affecting the behavior of the phase diagram of LiHoF$_4$ in the 
 $B_x/T_c\rightarrow0$ regime.
 We derive 
the required effective model 
 in two steps. Firstly, in LiHoF$_4$, in the temperature 
 range that we are interested in, which is close or below $T_c(B_x=0)=1.53$~K, the high energy 
 scales are well separated from the low energy sector. The energy scale for dipolar interactions
  between nearest-neighbor  Ho$^{3+}$ ions is  
   about 0.31~K. This is much smaller
 than the energy gap
 between the two first lowest single ion energy states and the next higher crystal field
 states ($>11$~K). In this case, one can neglect the higher energy states and reduce the full
 Hamiltonian Hilbert space to a smaller subspace spanned by the two lowest energy states. 
 This enables 
 us to deduce a low energy effective spin-$\frac{1}{2}$ Hamiltonian for LiHoF$_4$. Secondly, 
 we derive a semi-classical effective Hamiltonian from this low energy spin-$\frac{1}{2}$
  Hamiltonian 
 by incorporating the transverse field term perturbatively via a cumulant expansion. 
 We can then
perform a simple
classical
 Monte Carlo using this semi-classical effective Hamiltonian to investigate the small
  $B_x/T_c$ regime.   
      
\subsection{Effective Spin-$\frac{1}{2}$ Hamiltonian}

As mentioned in the previous section, there are three type of interactions that play a
 role in the magnetic properties of
LiHoF$_4$. The main interaction is the long-range dipole-dipole interaction between
 the Ho$^{3+}$ magnetic ions denoted by 
 \begin{eqnarray}
 H_{\rm dip}=\frac{1}{2}(g_{\rm L}\mu_{\rm B})^2\sum_{i\ne
j}\sum_{\mu\nu}L_{ij}^{\mu\nu}\rm{J}^{\mu}_i\rm{J}^{\nu}_j~,
\end{eqnarray}
 where $\mu,\nu$=$x,y,z$ and $\mathbf{J}_i$ is the total angular momentum of Ho$^{3+}$ ion 
 $i$. $L_{ij}^{\mu\nu}$ is the magnetic dipole interaction written in the form
$L_{ij}^{\mu\nu}=\left[ \delta^{\mu\nu}|
{\bf r}_{ij}|^2 -
3({\bf r}_{ij})^{\mu}({\bf r}_{ij})^{\nu}\right]/|{\bf r}_{ij}|^5$, where ${\bf r}_{ij}$
is the distance between ion $i$ and $j$.
$g_{\rm L}=1.25$ 
is the Land\'e g-factor of free Ho$^{3+}$ 
and $\mu_{\rm B}=0.6717~{\rm K}/{\rm T}$ is the Bohr magneton.
 The dipolar interaction is complemented by a short range 
 nearest-neighbor Heisenberg exchange interaction 
 \begin{eqnarray}
 H_{\rm exch}&=&
 \frac{1}{2}J_{\rm{ex}}\sum_{i,{\rm \rm NN}}\mathbf{J}_i\cdot
\mathbf{J}_{\rm \rm NN}~, 
\end{eqnarray}
where ${\rm NN}$ denotes the
 nearest neighbors of site $i$. 
This exchange interaction is considered to be weak and isotropic~\cite{prabuddha,LiTb}. 
The third interaction is the hyperfine coupling 
between the electronic and nuclear magnetic moments 
\begin{eqnarray}
H_{\rm hyp}=A\sum_i(\mathbf{I}_i\cdot\mathbf{J}_i)~.
\end{eqnarray}
The hyperfine constant $A\approx 39$~mK is anomalously large in Ho$^{3+}$-based materials~\cite{Bitko,prabuddha,Schechter}. 
  Thus, the complete Hamiltonian  is written as 
 \begin{eqnarray}
  H &=& \sum_iV_C(\mathbf{J}_i) - g_{\rm L}\mu_{\rm B}\sum_iB_{x}{\rm{J}}_i^x \nonumber\\
  && + H_{\rm dip}+H_{\rm exch}+H_{\rm hyp}~.
\label{origham}
 \end{eqnarray}
 The first two terms are single ion interactions,
 where $V_C$ describes the strong crystal field interactions 
  discussed in  Section~\ref{Sec-2} and Appendix \ref{A}. The second term is the Zeeman interaction.
%
Henceforth, we ignore $H_{\rm hyp}$ since our goal, as explained in the Introduction, is to investigate the small 
$B_x$ and small $\left( T_c(0)-T_c(B_x)\right)/T_c(0)$ regime where, as already suggested by the results of
Ref.~[\onlinecite{prabuddha}] and shown in Fig.~\ref{whole}, the hyperfine interaction effects are negligible.
The first two single-site (non-interacting) terms in $H$, denoted as
\begin{equation}
H_{\rm single-site}=V_{C}(\mathbf{J}) - g_{\rm L}\mu_{\rm B}B_{x}\rm{J}^{x}~,
\label{truncate}
\end{equation}
can be easily numerically diagonalized  for arbitrary transverse
field $B_x$ \cite{prabuddha}. $|\alpha(B_{x})\rangle$ and $|\beta(B_{x})\rangle$ are 
the two lowest states of the single ion Hamiltonian~(\ref{truncate})
 for a given $B_{x}$. Their corresponding energies are denoted 
 by $E_{\alpha}(B_{x})$ and $E_{\beta}(B_{x})$.
\begin{figure}[t]
\includegraphics[width=\columnwidth,angle=0]{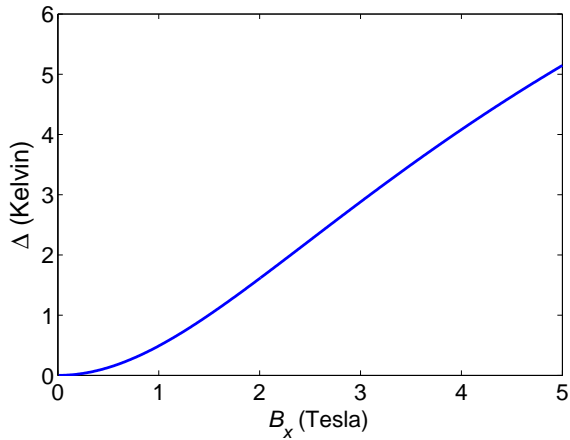}
\caption{\small{ The energy splitting of the ground state doublet, 
$\Delta(B_{x})\equiv E_{\beta}(B_{x}) -
E_{\alpha}(B_{x})$, 
in LiHoF$_4$ as a function of $B_x$ the transverse magnetic field.
The crystal field $V_c$ was obtained from Refs.~[\onlinecite{prabuddha,Ronnow}].
For more details on the crystal field and crystal field parametrization
(see Appendix \ref{A}).
\label{Delta}}}
\end{figure}
 At $B_x=0$ these two states form a doublet, but $B_x \neq 0$ lifts the degeneracy.  
 The Ising subspace $|\! \uparrow\rangle$ and $|\! \downarrow\rangle$
 are chosen by performing a unitary rotation on the $|\alpha(B_{x})\rangle$ and $|\beta(B_{x})\rangle$
 states~:
\begin{eqnarray}
|\! \uparrow\rangle &=& \frac{1}{\sqrt{2}}(|\alpha\rangle +
\exp(i\theta)|\beta\rangle)\nonumber\\
|\! \downarrow\rangle &=&
\frac{1}{\sqrt{2}}(|\alpha\rangle - \exp(i\theta)|\beta\rangle)	\; .
\label{updown}
\end{eqnarray} 
The phase $\theta$ is chosen such that the matrix elements of the
operator J$^{z}$ between $|\! \uparrow\rangle$ and $|\! \downarrow\rangle$
is real and diagonal, giving for J$_i^{z}$,
$
{\rm{J}}^{z}_{i}=C_{zz}\sigma^{z}_{i}. \label{zmap1}
$
Since the first excited state, $|\gamma(B_{x})\rangle$, above $|\alpha(B_{x})\rangle$ and $|\beta(B_{x})\rangle$,
is at an energy at least seven times higher than $k_BT_c(B_x)$, and is repelled for all $B_x$
from the $|\alpha(B_{x})\rangle$ and $|\beta(B_{x})\rangle$ set (see Fig.~1 of
Ref.~[\onlinecite{prabuddha}]), we henceforth neglect all excited crystal field states and work in a reduced 
Hilbert space spanned solely
by $|\alpha(B_{x})\rangle$ and $|\beta(B_{x})\rangle$,
or equivalently by  $|\! \uparrow\rangle$ and $|\! \downarrow\rangle$.
Projecting the single ion Hamiltonian of Eq.~(\ref{truncate}) in this two-dimensional 
subspace for an arbitrary ion $i$, we get 
\begin{equation}
H_{T}=\overline{E}(B_{x}) - \frac{1}{2}\Delta(B_{x})\sigma^{x},
\label{splitting}
\end{equation}
where $\overline{E}(B_{x})=\frac{1}{2}(E_{\alpha}(B_{x}) +
E_{\beta}(B_{x}))$ and $\Delta(B_{x})=E_{\beta}(B_{x}) -
E_{\alpha}(B_{x})$.  The energy difference
between the two lowest states caused by the transverse magnetic field $B_x$ can
already be interpreted as an effective transverse field 
$\Gamma={\Delta(B_x)}/{2}$ acting on
$S_{\rm{eff}}$=$\frac{1}{2}$ degrees of freedom at each site.
The dependence of $\Delta(B_{x})$ on the magnetic transverse field $B_x$
 is plotted in Fig.~\ref{Delta}.

Since we are henceforth working in a two-dimensional subspace for each ion $i$, we can write
the interactions between J$_i^\mu$ and J$_j^\nu$ in terms of 
effective interactions between Pauli matrices.                            
Indeed, any operator acting
in a two-dimensional space can be written as a linear combination 
of $\sigma_i^{\mu}$ Pauli matrices plus the unit matrix $\sigma^0\equiv \openone$.        
In order to express J$_i^\mu$ in terms of $\sigma_i^{\mu}$, we project J$_i^\mu$
in the subspace spanned by $|\! \uparrow\rangle$ and $|\! \downarrow\rangle$. 
\begin{figure}[bp]
\includegraphics[width=8cm,angle=0]{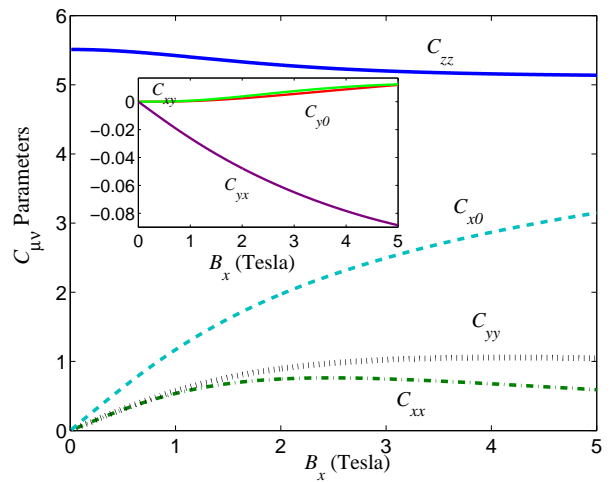}
\caption{\small{The evolution of the $C_{\mu\nu}$ parameters using the crystal field $V_c$ from 
Refs.~[\onlinecite{prabuddha, Ronnow}]. In the inset one can see that $C_{xy}\approx C_{y0}$.
Coefficients that are not plotted are zero.}}
\label{evolution}
\end{figure}
Specifically, we write the J$^{\mu}$ operator as
\begin{equation}
{\rm{J}}^{\mu}=C_{\mu 0}\sigma^{0} + \sum_{\nu=x,y,z}C_{\mu\nu}(B_{x})\sigma^{\nu}~,
\label{map1}
\end{equation}
where 
\begin{eqnarray*}
C_{\mu z}&=&\frac{1}{2}\left[ \langle\uparrow \! |{\rm J}^{\mu}|\! \uparrow\rangle-
\langle\downarrow \!|{\rm J}^{\mu}|\! \downarrow\rangle\right]~,\\
C_{\mu 0}&=&\frac{1}{2}\left[ \langle\uparrow \!|{\rm J}^{\mu}|\! \uparrow\rangle+
\langle\downarrow \! |{\rm J}^{\mu}|\! \downarrow\rangle\right]~,\\
C_{\mu x}&=&\frac{1}{2}\left[ \langle\uparrow \! |{\rm J}^{\mu}|\!\downarrow\rangle+
\langle\downarrow \!|{\rm J}^{\mu}| \!\uparrow\rangle\right]~~~~{\rm and}\\
C_{\mu y}&=&\frac{1}{2i}\left[ \langle\uparrow \! |{\rm J}^{\mu}|\! \downarrow\rangle-
\langle\downarrow \! |{\rm J}^{\mu}|\! \uparrow\rangle\right]~.
\end{eqnarray*} 

Based on the crystal field parameters of Refs.~[\onlinecite{prabuddha, Ronnow}], 
the evolution of
 the various parameters $C_{\mu\nu}$ and $C_{\mu 0}$ as a function of $B_x$ 
 is plotted in Fig.~\ref{evolution}.
We see that
 $C_{zz}$ is the largest
  term compared to all the other $C_{\mu\nu}$'s.

\begin{figure}[tbp]
\center
\psfrag{a1}[c]{\begin{footnotesize}$\vert\frac{(g_{\rm L}\mu_{\rm B})^2}{\Delta}C_{x0}C_{xx}\sum_j L_{ij}^{xx}\vert$\end{footnotesize}}
\psfrag{b1}[c]{\begin{footnotesize}$\vert\frac{(g_{\rm L}\mu_{\rm B})^2}{\Delta}C_{yy}^2\sum_j L_{ij}^{yy}\vert$\end{footnotesize}}
\psfrag{data3}[c]{\begin{footnotesize}$\vert\frac{(g_{\rm L}\mu_{\rm B})^2}{\Delta}C_{xx}^2\sum_j L_{ij}^{xx}\vert$\end{footnotesize}}
\psfrag{x1}[c]{$B_x$~(Tesla)}
\psfrag{y1}[c]{Ratio of Coefficients}
\includegraphics[width=93 mm,angle=0]{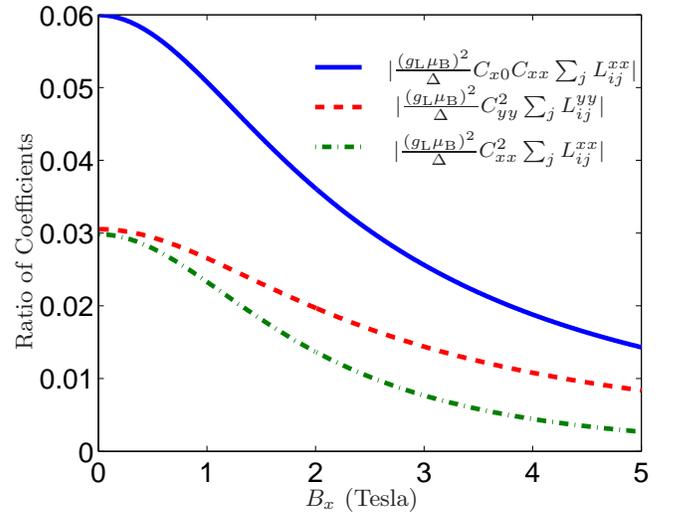}
\caption{\small{The ratio of the typical value of terms neglected in 
Hamiltonian~(\ref{isingdip}) respect to $\Delta$, using the crystal field $V_c$ from 
Refs.~[\onlinecite{prabuddha, Ronnow}] and the dipolar sum is performed for a long cylindrical sample. }}
\label{ratio}
\end{figure}

For the Hamiltonian in Eq.~(\ref{origham}), 
the J$_i^\mu$ operators are substituted by their two dimensional  
representations introduced in 
Eq.~(\ref{map1}). 
 This leads to a complicated Hamiltonian
that acts within the Ising subspace of $|\! \uparrow\rangle$ and $|\! \downarrow\rangle$.
 The projection generates various
kinds of interactions among the effective S$_{\rm eff}=\frac{1}{2}$ spins.
Via Eq.~(\ref{updown}), a specific rotated subspace was chosen, such that 
$C_{z\mu}=0$ ($\mu=x,y,0$; $\sigma^0\equiv \openone$).
 As shown in the inset of Fig.~\ref{evolution}, $C_{xy}$, $C_{yx}$, and $C_{y0}$
are very small, so the interacting terms containing these coefficients can be neglected. 
Therefore, neglecting these terms, we obtain
\begin{widetext}
\begin{eqnarray}
H_{\rm spin-1/2}&=&\frac{1}{2}(g_{\rm L}\mu_{\rm B})^2\left[\right.C_{zz}^2(B_x)\sum_{i\ne
j}L_{ij}^{zz}\sigma^{z}_i\sigma^{z}_j 
+2C_{zz}(B_x)C_{xx}(B_x)\sum_{i\ne j}L_{ij}^{zx}\sigma^{z}_i\sigma^{x}_j
\nonumber\\
& &+2C_{zz}(B_x)C_{yy}(B_x)\sum_{i\ne j}L_{ij}^{zy}\sigma^{z}_i\sigma^{y}_j + C_{xx}^2(B_x)\sum_{i\ne j}L_{ij}^{xx}\sigma^{x}_i\sigma^{x}_j
+ C_{yy}^2(B_x)\sum_{i\ne j}L_{ij}^{yy}\sigma^{y}_i \sigma^{y}_j\left.\right]\nonumber\\ 
& & + \frac{1}{2}{\rm J_{\rm ex}}\sum_{\mu}C_{\mu\mu}^2(B_x)\sum_{i,{\rm \rm NN}}
\sigma^{\mu}_i\sigma^{\mu}_{\rm \rm NN}
+(g_{\rm L}\mu_{\rm B})^2C_{zz}(B_x)C_{x0}(B_x)\sum_{i\ne j}L_{ij}^{zx}\sigma^{z}_i
\nonumber\\
& &  
+ \sum_i \left[C_{x0}(B_x)C_{xx}(B_x)\left(4J_{\rm ex}+(g_{\rm L}\mu_{\rm B})^2\sum_j L_{ij}^{xx}\right)
-\frac{\Delta(B_x)}{2}\right]\sigma^x_i~. \label{full2D}
\end{eqnarray}
\end{widetext}
When the external magnetic field $B_x$ is zero, 
only $C_{zz}(0)\neq 0$ 
 and all the other $C_{\mu\nu}$ and $C_{\mu 0}$ vanish.
 Hence, in absence of an external magnetic field, the system can 
 be described by a simple classical dipolar Ising model~\cite{prabuddha}. 
Fortunately, a number of interaction terms are zero or can be neglected with respect to the 
leading Ising interaction, which is proportional to
 $C_{zz}^2(B_x)\sum_{i\ne j}L_{ij}^{zz}\sigma^{z}_i\sigma^{z}_j$.
 As we can see from Eq.~(\ref{full2D}), for pure LiHoF$_4$, an effective 
   $\sigma_i^x\sigma_j^x$ and $\sigma_i^y\sigma_j^y$ pair-wise interactions 
as well as a linear transverse field along
the $x$ direction are induced in the presence of an
    external magnetic field.
  As suggested by Fig.~\ref{ratio}, and already assumed in Ref.~[\onlinecite{prabuddha}],
   we expect the quantum fluctuations 
    induced by these terms via either dipolar or exchange coupling, to be quite small and negligible compared to the quantum fluctuations
    induced by $\Delta(B_x)$.
For the pure (disorder free) LiHoF$_4$, 
the invariance of the dipolar interactions under lattice mirror symmetries forces 
$\sum_{
j}L_{ij}^{zx}=0$. 
 So the linear term with $C_{zz}(B_x)
C_{x0}(B_x)\sum_{i\ne
j}L_{ij}^{zx}\sigma^{z}_i$ vanishes.
Considering the $C_{zz}(B_x)
C_{xx}(B_x)\sum_{i\ne j}L_{ij}^{zx}\sigma^{z}_i\sigma^{x}_j$ term, 
because of lattice mirror symmetry, one has 
$\sum_{i\ne j}L_{ij}^{zx}\sigma^{z}_i\left\langle \sigma^{x}_j\right\rangle=0$,
therefore this term can only contribute via 
high order fluctuation effects beyond the
vanishing mean-field contribution.
Since $\frac{C_{x0}(B_x)}{C_{zz}(B_x)}<1$, we expect the (second order) 
fluctuation contribution effects from the above $\sigma^{z}_i\sigma^{x}_j$ term to be small.
Hence we neglect
 the  $C_{zz}(B_x)
C_{xx}(B_x)\sum_{i\ne j}L_{ij}^{zx}\sigma^{z}_i\sigma^{x}_j$ term in
the S$_{\rm eff}=\frac{1}{2}$ effective Hamiltonian $H_{\rm spin-1/2}$.
We should emphasize that for diluted LiHo$_x$Y$_{1-x
}$F$_4$, since the lattice mirror symmetries are broken,
 the two latter terms, proportional to $\sum_{i\ne
j}L_{ij}^{zx}\sigma^{z}_i$ and 
$\sum_{i\ne j}L_{ij}^{zx}\sigma^{z}_i\left\langle \sigma^{x}_j\right\rangle$, 
can no longer be neglected~\cite{Tabei-PRL}.
Indeed, these are the terms responsible for the generation of the 
longitudinal random fields
in LiHo$_x$Y$_{1-x}$F$_4$ when subject to nonzero $B_x$~\cite{Schechter-PRL2,Tabei-PRL,Tabei-Vernay},
as discussed
in the Introduction.


  Hence,
    the spin-$\frac{1}{2}$ Hamiltonian in Eq.~(\ref{full2D})
    can be further simplified to a familiar looking transverse field
Ising Hamiltonian with a dipolar
    and nearest-neighbor exchange Ising interaction. 
\begin{eqnarray}
H_{\rm spin-1/2}&=&\frac{1}{2}C_{zz}^2(B_x)\left[(g_{\rm L}\mu_{\rm B})^2\sum_{i\ne
j}L_{ij}^{zz}\sigma^{z}_i\sigma^{z}_j\right. \nonumber\\
&+&\left. J_{\rm ex}\sum_{i,\rm NN}
\sigma^{z}_i\sigma^{z}_{\rm NN}\right]
-\frac{\Delta(B_x)}{2}\sum_i 
\sigma^x_i. \label{isingdip}
\end{eqnarray} 
To simplify the calculations, and in order to be consistent with the notation of
Ref.~[\onlinecite{prabuddha}] as well as for further comparison between our simulation
 results and those of Ref.~[\onlinecite{prabuddha}],
 we lump the whole $B_x$ dependence
 in the transverse field term into a
  renormalization factor $\epsilon(B_x)$ is defined as
\begin{eqnarray}
\epsilon(B_x)=\frac{C_{zz}(B_x)}{C_{zz}(0)}~.
\label{epsilon}
\end{eqnarray}
We renormalize the Hamiltonian as
\begin{eqnarray}
H_{\rm spin-1/2}=\left[ \epsilon(B_x)\right]^2\widetilde{\mathcal{H}}~,
\end{eqnarray}
with, according to Eq.~(\ref{isingdip}), $\widetilde{\mathcal{H}}$ is
\begin{eqnarray}
\widetilde{\mathcal{H}}&=&\frac{1}{2}C_{zz}^2(0)\left[(g_{\rm L}\mu_{\rm B})^2\sum_{i\ne
j}L_{ij}^{zz}\sigma^{z}_i\sigma^{z}_j 
+J_{\rm ex}\sum_{i,\rm NN}
\sigma^{z}_i\sigma^{z}_{\rm NN}\right]\nonumber\\
&&- g_{\rm L}\mu_{\rm B} C_{zz}(0)\widetilde{\mathcal{B}}_x\sum_i 
\sigma^x_i~, \label{Risingdip}
\end{eqnarray}
where the renormalized effective transverse magnetic field
$\widetilde{\mathcal{B}}_x$, is related to the real applied $B_x$ via
\begin{eqnarray}
\widetilde{\mathcal{B}}_x=\frac{\Delta(B_x)}{2g_{\rm L}\mu_{\rm B} C_{zz}(0)\times
\left[ \epsilon(B_x)\right]^2}	,
\label{effectiveB}
\end{eqnarray}
consistent with Ref.~[\onlinecite{prabuddha}]. In discussing 
Monte Carlo simulations below,
we also define a renormalized temperature, $\widetilde{T}$, in conjunction with 
$\widetilde{\mathcal{H}}$, with
$\widetilde{T}$ 
defined as
\begin{eqnarray}
T=\left[ \epsilon(B_x)\right]^2\widetilde{T},
\label{effectiveT}
\end{eqnarray}
where $T$ is the real physical temperature.

All results presented in the Monte Carlo simulations section below were obtained by considering
the renormalized
Hamiltonian~(\ref{Risingdip}), and performing the simulations  with respect to the 
renormalized $\widetilde{T}$ and 
$\widetilde{\mathcal{B}}_x$. Before presenting our  Monte Carlo simulations of Eq.~(\ref{Risingdip})
as pertain to LiHoF$_4$, we first discuss the technique we 
employed to handle
quantum fluctuations
 perturbatively for
small $B_x/T_c$.

\subsection{Effective classical temperature-dependent Hamiltonian $-$ perturbation expansion}

In this section, with a focus on the simplified spin $\frac{1}{2}$
 Hamiltonian of Eq.~(\ref{Risingdip}), 
 we aim to implement a cumulant perturbative Monte Carlo method 
 for a spin $\frac{1}{2}$ transverse Ising model~\cite{precumulant,cumulant}. 
  For small quantum fluctuations, close to the classical
critical temperature, we are able to derive an effective classical
 Hamiltonian analytically, where quantum fluctuations are incorporated perturbatively.
 Using such effective perturbative Hamiltonian, we can then perform classical MC simulations.   
  To set the stage, we first consider a general transverse field Ising Hamiltonian such as
\begin{eqnarray}
\mathcal{H}&=&\frac{1}{2}\sum_{i,j}\mathcal{L}_{ij}^{zz}\sigma_i^z\sigma_j^z +\frac{1}{2}
\mathcal
{J}_{\rm ex}\sum_{i,\rm NN}\sigma_i^z\sigma_{\rm NN}^z\nonumber\\
& & -\Gamma\sum_i\sigma_i^x
-h_0\sum_i\sigma_i^z.
\label{HOneHalf}
\end{eqnarray} 
$\Gamma$ is the transverse field in the $x$ direction and $h_0$ denotes an
external longitudinal field along the $z$ direction.
For compactness, note that we passed from dipolar interactions denoted 
$C_{zz}^2(0)(g_{\rm L}\mu_{\rm B})^2L_{ij}^{zz}$ to $\mathcal{L}_{ij}^{zz}$ and
from exchange interaction
$C_{zz}^2(0)J_{\rm ex}$ to $\mathcal{J}_{\rm ex}$
(
see Eq.~(\ref{Risingdip})
). 
 The partition function $Z$ for a system with Hamiltonian (\ref{HOneHalf}) is 
\begin{eqnarray}
Z&=&{\rm Trace}(e^{-\beta\mathcal{H}})\nonumber\\
&=&\sum_{\left\lbrace \psi_i\right\rbrace}\langle \psi_i\vert e^{-\beta \mathcal{H}}\vert\psi_i\rangle,
\end{eqnarray} 
where $Z$ is obtained by tracing over ${\psi_i}$'s which are, for example, direct product 
 of $\sigma_i^z$ 
eigenvectors~($| \! \uparrow\rangle$ and $| \! \downarrow \rangle$) and
 $\beta\equiv1/k_{\rm B}T$. 
We can write the Hamiltonian (\ref{HOneHalf}) as $\mathcal{H}=H_0+H_1$. $H_0$ is 
the classical part
 of the Hamiltonian, for which the ${\psi_i}$'s are eigenvectors.  
$H_1 \equiv -\Gamma\sum_i\sigma_i^x$ is the quantum term, which does not commute with $H_0$.
The existence of these two non-commuting terms in $\mathcal{H}$ 
prevents us from applying classical Monte Carlo
 techniques directly to the system. We can derive an effective classical Hamiltonian as
 a functional of ${\psi_i}$, such that
\begin{eqnarray}
e^{-\beta H_{\rm eff} \left[ \psi_i\right]} =\langle \psi_i\vert e^{-\beta \mathcal{H}}\vert\psi_i\rangle~ .
\label{matrixelement}
\end{eqnarray} 
 Referring to the definition above in Eq.~(\ref{matrixelement}), since the right hand side of Eq.~(\ref{matrixelement})
 is the matrix element with respect to $\vert\psi_i\rangle$, 
 $H_{\rm eff}\left[ \psi_i\right]$ is a functional depending only on the set of $\sigma_i^z$ 
eigenvalues. 
The partition function can then be written as a classical partition function  
\begin{eqnarray}\label{classical_part}
Z=\sum_{\left\lbrace \psi_i\right\rbrace}e^{-\beta H_{\rm eff}\left[\psi_i \right] }~.
\end{eqnarray} 
By finding an explicit expression for $H_{\rm eff}\left[\psi_i \right]$, one can calculate
 the thermodynamical properties of the system described by ${\mathcal H}$
by performing {\it classical} Monte Carlo simulations using $H_{\rm eff}$ instead of 
${\mathcal H}$.\\

To proceed, we write
the matrix element $\langle \psi\vert e^{-\beta \mathcal{H}}\vert\psi\rangle$ in terms of a cumulant
 expansion~\cite{cum}
\begin{eqnarray}\label{cumulant}
\langle \psi\vert e^{-\beta\mathcal{H}}\vert\psi\rangle = \hspace{65 mm}\nonumber\\
\exp\left[-\beta \langle\psi|\mathcal{H}|\psi \rangle
 +  
\sum_{n>1}^{\infty}\frac{\left( -\beta\right)^n}{n!}
\langle\psi\vert\left( \mathcal{H}-\langle\psi|\mathcal{H}|\psi\rangle\right)^n\vert\psi\rangle\right]~.
\end{eqnarray}
To make the notation more compact, by  $\vert\psi\rangle$ we mean a
typical  $\vert\psi_i\rangle$ eigenvector.
Using Eq.~(\ref{cumulant}) we can derive the effective Hamiltonian
 $H_{\rm eff}\left[ \psi_i\right]$ perturbatively. 
 The details of the derivation of $H_{\rm eff}\left[ \psi_i\right]$
 are presented in Appendix~\ref{B}.  $H_{\rm eff}\left[ \psi_i\right]$,
  is to order $O(\Gamma^2)$, given by 
\begin{eqnarray}
H_{\rm eff}
&=& H_0+\beta\Gamma^2\sum_{i}\{\sigma_i^zF_1\left[2\beta(h_i+h_0)\right]\nonumber\\
&&- F_0\left[ 2\beta(h_i+h_0)\right]\}.
\label{eff} 
\end{eqnarray} 
  In Eq.~(\ref{eff}), $h_i$ is the total local field affecting the spin at site $i$ caused
  by all the other spins, and which is
\begin{eqnarray}
h_i=-\sum_{j\neq i}\mathcal{L}_{ij}^{zz}\sigma_j^z-\mathcal{J}_{\rm ex}\sum_{\rm NN}\sigma_{\rm \rm NN}^z~,
\label{localfield}
\end{eqnarray}
and $h_0$ is the external longitudinal field in the $z$ direction. 
The functions $F_0(x)$ and $F_1(x)$ are defined as
\begin{eqnarray}
F_0(x)	\equiv \frac{\cosh(x)-1}{x^2},\nonumber \\
F_1(x)	\equiv \frac{\sinh(x)-x}{x^2}.
\end{eqnarray}
~~In this effective Hamiltonian, the effect of quantum fluctuations is taken into
account perturbatively to order  $O(\beta\Gamma^2/[H_0])$, where $[H_0]$ denotes the order of magnitude of
$H_0$, the classical part
(first two terms)
of Eq.~(\ref{HOneHalf}). 
To obtain the thermodynamical
 properties of the system
 for small transverse fields we can therefore perform a classical Monte-Carlo on $H_{\rm eff}$ as a classical
  counterpart of the real quantum mechanical Hamiltonian.  
  Since we are interested in thermal averages we can 
 calculate thermodynamical quantities by differentiating 
 the partition function, which is written in terms of  $H_{\rm eff}\left[ \psi_i\right]$,
  with respect to $h_0$, $\Gamma$ or $\beta$. 
  The effective Hamiltonian has an explicit $h_0$ and $\beta$
 dependence.
For each true thermodynamical quantum-mechanical quantity, 
we obtain a pseudo-operator counterpart.
 For example the
 pseudo-operators corresponding
to $\langle E \rangle$, $\langle M_z \rangle$, $\langle M_x \rangle$,  $\langle M_z^2 \rangle$, and
 $\langle M_z^4 \rangle$ are calculated in Appendix~\ref{C}, where $E$, $M_z$ and $M_x$ are the energy and 
 magnetization operators along the $z$ and $x$ direction. $\langle\dots\rangle$ 
 stands for the Boltzmann thermal average.  
 
\begin{figure}[htp]
\includegraphics[width=\columnwidth,angle=0]{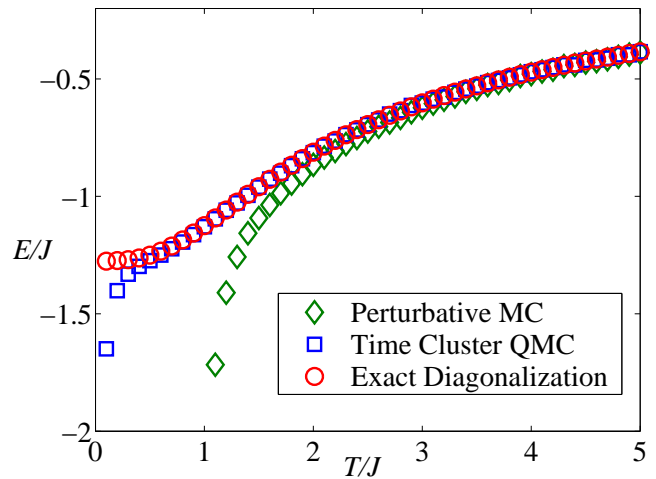}
\caption{Energy as a function of temperature for a simple one dimensional
nearest-neighbor Ising chain with a transverse field of $\Gamma=J$ and $N=10$ spins
and periodic boundary conditions. The energy is obtained by exact diagonalization 
of the Hamiltonian, a time-cluster QMC algorithm, and a classical 
Monte-Carlo algorithm of the 
perturbative effective Hamiltonian.\label{fig:1Dchain}}
\end{figure}

 Because of its perturbative nature in ($\beta\Gamma$), this method is not
 reliable for large transverse fields or 
 low temperatures. To illustrate the range of validity of this method we consider a simple
 one-dimensional nearest-neighbor transverse-field Ising-model Hamiltonian
  $H=-J\sum_i\sigma_i^z\sigma_{i+1}^z -\Gamma\sum_i\sigma_i^x$ with periodic 
  boundary conditions. 
  For a one-dimensional chain of 10 ions, we are able to calculate the exact total 
  energy of the chain by exact diagonalization.
 To check our perturbative MC technique, we calculated
  the energy of the Ising chain as a function of temperature for a given transverse field.
   To make a comparison, 
  we  also performed a quantum Monte-Carlo~(QMC) simulation on the system. In this QMC simulation, 
  we used the Trotter-Suzuki~\cite{trotter} formalism and 
   applied a continuous time cluster algorithm similar to the one in Ref.~[\onlinecite{rieger}].
   In Fig.~\ref{fig:1Dchain}, for a quite large transverse field $\Gamma/J=1$, we  plot
    the average thermal energy as a function of temperature obtained from exact diagonalization,
     time cluster QM and ``perturbative  MC" using the effective perturbative
      Hamiltonian described above.
This tests confirms the
quantitative correctness of the perturbative Monte Carlo scheme
 at small $\beta\Gamma^2/J$.
      We also computed other thermodynamic quantities 
      (e.g. $\langle M_z \rangle$, $\langle M_x \rangle$ ) and these also compared well
      with QMC and exact diagonalization results.

Before we present our Monte Carlo results for LiHoF$_4$, let us summarize what we have done so far.
\begin{enumerate}
\item Since the spin-spin interactions and $T_c(B_x)$ are small compared to the gap between the low-lying states
$|\alpha(B_x)\rangle$ and $|\beta(B_x)\rangle$ with respect to the excited
crystal field  state
 $|\gamma(B_x)\rangle$, we can recast the full microscopic model of LiHoF$_4$ in terms of an effective 
 transverse field Ising model with effective spin-spin interactions and effective transverse field $\Gamma(B_x)$
 that depend on the real physical applied magnetic field $B_x$.
\item Since we are interested in a regime where $B_x/T_c$ is small, we can develop a perturbation expansion of 
the partition function in powers of $B_x/T$ and  recast the thermal averages of real physical observables in terms of quantities 
that can be determined via a classical Monte Carlo simulation of a further effective temperature-dependent classical Hamiltonian. 
\end{enumerate}
Having shown that the perturbative cumulant MC can quantitatively describe the TFIM for
small $\beta\Gamma^2/[H_0]$, we proceed in the next section to describe how we use this method to study LiHoF$_4$
at small transverse field $B_x$, $B_x/T_c \ll 1$.

\section{Perturbative Monte Carlo Study of ${\rm{\bf LiHoF}}_4$}

In this section we report results from the perturbative Monte Carlo (MC) simulation to study 
the low transverse field $B_x$ properties 
of LiHoF$_4$, using the low field perturbative effective Hamiltonian in Eq.~(\ref{eff}) 
and using Eq.~(\ref{localfield}) for the definition of the local $h_i$ fields. 
As discussed in the Introduction, our primary goal here is to check 
the quantum
Monte Carlo results from stochastic series expansion of Ref.~[\onlinecite{prabuddha}],
 and investigate
the contrasting 
results with the transverse field $B_x$ phase diagram of Ref.~[\onlinecite{Bitko}] 
for small $B_x$ (See Fig.~\ref{whole}).
Hence, we are indeed interested in LiHoF$_4$ in the case of asymptotically small
$B_x/T_c$.
The temperature we use in our simulations is the renormalized 
temperature defined in Eq.~(\ref{effectiveT}). 
Regarding Eq.~(\ref{Risingdip}), the transverse field $\Gamma$ used in the 
perturbative effective Hamiltonian~(\ref{eff}) is 
$\Gamma=g_{\rm L}\mu_{\rm B} C_{zz}(0)\widetilde{\mathcal{B}}_x$,
where $\widetilde{\mathcal{B}}_x$ is defined in Eq.~(\ref{effectiveB}).
For the local field $h_i$, defined in Eq.~(\ref{localfield}), we have 
$\mathcal{L}_{ij}^{zz}=C_{zz}^2(0)(g_{\rm L}\mu_{\rm B})^2L_{ij}^{zz}$
and
$ 
\mathcal{J_{\rm ex}}=C_{zz}^2(0)J_{\rm ex}
$~.

In the following subsections, we first discuss the 
reaction field (RF) and the Ewald summation (ES) methods
that we use to deal with the long 
range dipolar interactions, and discuss how the Monte Carlo results in the classical regime,
where $B_x=0$, are affected by 
the choice of the method we use. 
Next, we discuss the sensitivity of the $J_{\rm ex}$ estimates
at zero $B_x$
to finite-size effects, boundary conditions and choice of the method
to handle the dipolar lattice sum.
 We also consider the effect of
 different $J_{\rm ex}$ on the phase digram, when $B_x \neq 0$ and 
 $B_x/T$ is small. Finally, we investigate to what extent the final results depend
 on the set of crystal field parameters chosen to describe the Ho$^{3+}$
 single ion properties.

\subsection{Reaction Field Method vs Ewald Summation Method}

Griffiths' theorem~\cite{Griffiths2} states that 
  in the absence of an external field
  the free energy for a dipolar lattice system has to be independent of the sample shape
  in the thermodynamical limit.  Therefore, as an immediate consequence,   
  in the absence of an external field, the net magnetization $\mathcal{M}$ of the sample 
  has to be zero. Otherwise, for a
uniform $\mathcal{M} \neq 0$, a shape dependent demagnetization field would 
    couple to the dipolar moments of the sample, making the free energy shape dependent.
    Here, the demagnetization field is the field originating from the magnetic moments sitting on 
   the boundary of the sample. 
Hence,
  in the thermodynamic limit, domains form in order for the system to
have a zero magnetization, $\mathcal{M}=0$.
   
Experiments on LiHoF$_4$ show that the results 
are shape independent, confirming Griffiths theorem and domain formation~\cite{cooke,Battison}.
 There is evidence that in LiHoF$_4$
long needle-shaped domains form along 
the $c$ axis~\cite{cooke,Battison}.
If we assume that there is a uniform 
macroscopic bulk magnetization $\mathcal{M}_z$ within a long needle-shaped domain
and the external magnetic field acting on the domain is $B_z^{\rm ext}$, 
then the susceptibility $\chi$
 of the domain is
\begin{eqnarray}
\chi=\mathcal{M}_z/B_z^{\rm ext}~.
\end{eqnarray} 
It should be noted that the macroscopic bulk magnetization $\mathcal{M}_z$, is 
given by $\mathcal{M}_z=n_0 g_{\rm L}\mu_{\rm B}\left\langle J^z\right\rangle$~, where 
$n_0=4/a^2c$ is the number of dipoles per unit of volume and where 
 $a^2c$ is the volume of the unit cell. Using ${\rm{J}}^{z}=C_{zz}\sigma^{z}$, the bulk magnetization
$\mathcal{M}_z$  is related to the 
total moment of the effective Ising spins, $M_z=\sum_i \sigma_i^z$,
in the $S_{\rm eff}$=$1/2$ picture by
\begin{eqnarray}
\mathcal{M}_z=\frac{4}{N}\frac{g_{\rm L}\mu_{\rm B}C_{zz}(B_x)}{a^2c}\left\langle M_z\right\rangle~,
\end{eqnarray}
where $N$  is the total number of dipoles.

Let us consider
consider an imaginary \textit{macroscopic} spherical cavity  deep inside a
 needle-shaped domain. The magnetization inside the sphere should be equal
 to the uniform bulk magnetization of the long needle-shaped domain.
  Apart from the external magnetic field $B_z^{\rm ext}$, spins
 enclosed in the sphere experience an  additional field
 that originates from the spins on the outer boundary of the
 imaginary sphere embedded in the long needle-shaped domain.
 The magnetic surface charge density on the surface of the
needle-shaped domain with uniform magnetization 
 $\mathcal{M}_z$ produces an internal 
magnetic field $B_{\rm needle}= 4\pi\mathcal{M}_z$.
Meanwhile, the magnetic surface charge density
  on the surface of the uniformly magnetized sphere with 
magnetization of $\mathcal{M}_z$ induces a (demagnetization)
   magnetic field $\frac{8\pi}{3}\mathcal{M}_z$ inside the sphere
that is in the opposite direction to the applied field and to
$B_{\rm needle}$.
Therefore, the total field $B_z^{\rm sph}$ inside the  spherical 
cavity 
is~\cite{Jackson}
 \begin{eqnarray}
B_z^{\rm sph}=B_z^{\rm ext}-\frac{8\pi}{3}\mathcal{M}_z+4\pi \mathcal{M}_z.
\label{demagnet}
\end{eqnarray}
 $\mathcal{M}_z$
is uniform for a bulk sample. 
Now, instead of considering a whole needle-shaped bulk, we
can also study an isolated spherical sample which
an effective $B_z^{\rm sph}$ field is applied to it.
 If we substitute $B_z^{\rm ext}$ with $\mathcal{M}_z/\chi$ and $B_z^{\rm sph}$
 with $\mathcal{M}_z/\chi_{\rm sph}$, where $\chi_{\rm sph}$
 is the susceptibility of the spherical domain,
 then we can write $\chi$ as a function of $\chi_{\rm sph}$
 \begin{equation}
\chi=\frac{\chi_{\rm sph}}{1-{\frac{4\pi}{3}}\chi_{\rm sph}} .
\label{chisphere}
\end{equation}
If $\chi_{\rm sph}$ is obtained 
via some calculation procedure 
for a spherical sample, one can use Eq.~(\ref{chisphere})
to determine the macroscopic susceptibility of the 
bulk sample within which the sphere is embedded.
Specifically, simulations can be
performed on a finite size sphere,
 and the effect of the macroscopic bulk surrounding the sphere is incorporated in
 a mean-field manner by considering an effective 
field $B_z^{\rm sph}$ interacting with the spins inside
 the spherical sample. 
Using this method, called 
the reaction field (RF) method, Chakraborty \textit{et al.} calculated
the finite size sphere susceptibility $\chi_{\rm sph}$ by using the stochastic series expansion
  quantum Monte-Carlo method \cite{prabuddha,sandvik}. 
They considered an $N$ spin system enclosed by a sphere, where
 the susceptibility of the sphere is obtained from the spin-spin correlation.
Referring to Eq.~(\ref{chisphere}), the 
paramagnetic to ferromagnetic transition (criticality) 
within the macroscopic long needle-shaped domain
 occurs at the temperature
for which $\chi_{\rm sph}=\frac{3}{4\pi}$ 
occurs for a spherical sample.
It should be noted that this criteria is 
derived for macroscopic systems in 
the thermodynamic limit. Therefore, 
as discussed in Ref.~[\onlinecite{Leeuw}], 
because of the fluctuation of magnetic moments on the
boundary of a finite size surface,
 quantities such as specific heat and susceptibility obtained via
 the RF method, are quite sensitive to finite size effects.

%

 The Ewald summation (ES) method \cite{Zeeman,huang,Leeuw}
is an alternative approach used to 
 obtain reliable quantitative results for describing real dipolar materials in a 
 periodic boundary condition (PBC)~\cite{spinice,Melko}.
 In the ES method, in order to treat long-range dipolar interactions with PBC,
 the system is modeled by replicating the simulation cell of linear size
  $L$ into a large array of image copies.  
  The ES method
  generates an 
  effective dipole-dipole interaction 
  $\sum_{\mu,\nu}L_{\rm eff}^{\mu\nu}({\bf r}_{ij}){\bm \mu}_i^{\mu}{\bm \mu}_j^{\nu}$
  between two arbitrary magnetic moments, ${\bm \mu}_i$ and 
   ${\bm \mu}_j$ within the simulation cell. Here, ${\bm \mu}_i=g_{\rm L}\mu_{\rm B}{\bf J}_i$,
   $\mu,\nu$=$x,y,z$, and 
${\bf r}_j-{\bf r}_i$ where ${\bf r}_i$ is the position of moment $i$.
  This is done by periodically replicating the simulation cell with a volume of $\Omega_0=L^3a^2c$ 
   and summing convergently the interactions between the real spins $i$ and $j$
    in the specified volume of the simulation cell of size $L$, 
   and all the periodically 
   repeated images of $j$ as
   \begin{equation}
 L_{\rm eff}^{\mu\nu}({\bf r}_{ij})
 =\sum_{\rm \bf n}L^{\mu\nu}({\bf r}_{ij}+{\rm \bf n})~,
 \end{equation}
   where ${\rm \bf n}=(n_xLa,n_yLa,n_zLc)$ with $n_x$, $n_y$, $n_z$ integers.
   $L^{\mu\nu}({\bf r}_{ij})=L^{\mu\nu}_{ij} \equiv \left[ \delta^{\mu\nu}|
{\bf r}_{ij}|^2 -
3({\bf r}_{ij})^{\mu}({\bf r}_{ij})^{\nu}\right]/|{\bf r}_{ij}|^5$ are dipolar couplings, 
which can be written in a more compact form as 
$L^{\mu\nu}({\bf r}_{ij})=\nabla_i^{\mu}\nabla_j^{\nu}|{\bf r}_{ij}|^{-1}$. Therefore
\begin{equation}
L_{\rm eff}^{\mu\nu}({\bf r}_{ij})
 =\nabla_i^{\mu}\nabla_j^{\nu}\sum_{\rm \bf n}|{\bf r}_{ij}+{\rm \bf n}|^{-1}~.
\label{Leffmunu-2}
\end {equation}
The sum $\sum_{\rm \bf n}|{\bf r}_{ij}+{\rm \bf n}|^{-1}$ is 
calculated using the Ewald method, such that
the sum contain a real space sum plus a reciprocal 
space sum minus a self term~\cite{Zeeman,huang,Leeuw}
 \begin{eqnarray}
\sum_{{\rm \bf n}}|{\bf r}_{ij}+{\bf n}|^{-1}
&=& \sum_{{\rm \bf n}}
\frac{{\rm erfc}(\kappa|{\bf r}_{ij}+{\rm \bf n}|)}{|{\bf r}_{ij}+{\rm \bf n}|}\nonumber\\
&& + \frac{ 1}{\pi \Omega_0}\sum_{{\bf k}\neq 0}\frac{4\pi^2}{k^2}e^{-k^2/4\kappa^2}
\cos\left({\bf k}\cdot {\bf r}_{ij} \right)\nonumber\\
&&-\frac{\kappa}{\sqrt{\pi}}\delta_{ij}~.   
\label{Ewald-formula}
\end{eqnarray}
Here ${\rm erfc}(x)=\left(2/\sqrt{\pi}\right)\times\int_{x}^{\infty}\exp{-t^2}dt$
and ${\bf k}$ denotes the reciprocal vectors of the simulation cell. The convergence factor $\kappa$
is chosen such that the real space sum and the reciprocal space sum converge 
about equally rapidly~\cite{Zeeman,huang,Leeuw}.
   The simulation cell and all its replicated 
images are embedded altogether in a continuous medium.
    Additionally, each spin experiences a demagnetization field,
    which is originating from the 
   magnetic moments on the boundary of the system~\cite{Leeuw}. 
   This boundary contribution depends on the shape of the boundary of the macroscopic 
   sample that we are interested in modeling. i.e. for a long needle-shaped sample 
   the demagnetization field correction to the ES
   representation of the dipole-dipole interactions is zero 
   ~\cite{Leeuw}. However,
   for a bulk spherical sample, the magnetic polarization of the magnetic moments on the boundary 
   of the sphere
    induces a demagnetization field proportional to the magnetization of the sample 
    ${\cal M}=\frac{1}{\Omega_0}\sum_i{\bm \mu}_i$, 
    which creates an additional
 effective field acting on the 
    the magnetic moments. The net effect results in
 an extra effective interaction 
   \begin{equation}\label{spher-bound} 
   \frac{4\pi}{2\mu'+1} \frac{{\bm \mu}_i\cdot{\bm \mu}_j}{\Omega_0}
   \end{equation}  
   between magnetic moments ${\bm \mu}_i$ and 
   ${\bm \mu}_j$ to be incorporated in the simulation~\cite{Leeuw}.
In practive, the term in Eq.~(\ref{spher-bound}) is merely added
to $L_{\rm eff}^{\mu\nu}({\bf r}_{ij})$ in Eq.~(\ref{Leffmunu-2}), which 
itself is calculated via the ES expression of Eq.~(\ref{Ewald-formula}).
    Here, $L$ is the linear 
system size, ${\bm \mu}_i=g_{\rm L}\mu_{\rm B}{\bf J}_i$, 
   and $\mu'$ is the magnetic permeability of the surrounding continuum. 
   For a sample surrounded
   by vacuum $\mu'=1$~\cite{vacuum_comment}. 
This interactions is added to the effective dipolar 
interaction between spins $i$ and $j$,
    derived by the ES technique~\cite{Melko}.
   
   As a result, within the ES method, each spin 
   interacts with all the ``real''
 spins in the specified simulation cell of linear size  
$L$, and with all its replicated periodic images. Therefore, one would expect
  the system to behave  more
 like a macroscopic system than in the RF method.
 However, there are still some 
  finite size effects due to the artifact of having a periodic sequence of cells
of finite
  size  $L$. 
   Once an effective dipole-dipole interaction between spins $i$ and $j$ 
   within the simulation cell has been derived via the ES technique, one can 
   perform Monte Carlo simulations using the standard Metropolis algorithm. 
   Xu \textit{et al.} \cite{Xu} used this ES technique to simulate 
   long-range dipolar Ising interactions for both the body-centered cubic (BCC) 
   and body centered
   tetragonal lattices in zero applied field. In a more recent work~\cite{Biltmo},
    the ES technique was implemented in
   a Monte Carlo simulation study of  LiHo$_x$Y$_{1-x}$F$_4$ in zero applied field.
   In the next subsection we discuss the results of MC simulations 
   using the cumulant perturbative method. In our simulations, we incorporate the long-range 
   dipolar interactions using both the RF method as discussed in Ref.~
   [\onlinecite{prabuddha}] and the ES method. The influence of each method on the MC 
   results is investigated in some detail.
\subsection{Perturbative Monte Carlo Simulations Results}


 In this subsection we describe the Monte-Carlo results obtained 
 using the effective perturbative Hamiltonian~(\ref{eff}) 
and which employ different ways to handle the dipolar lattice sums.
  We first report results obtained using
  the reaction field method for a spherical sample embedded in a long needle-shaped domain.
We also report  results from simulations using the ES
 method for both a long needle-shaped sample
   and a spherical sample embedded in a long 
  needle-shaped domain.

\subsubsection{Results from reaction field method}

\label{RF-method-results}

\begin{figure}[htp]
\center
\psfrag{T}[c]{$\widetilde{T}$~(K)}
\psfrag{chi}[c]{~~\large{$\chi_{\rm sph}$}}
\psfrag{pi}[c]{$\chi_{\rm sph}=\frac{3}{4\pi}$}
\psfrag{afdata}[c]{\begin{footnotesize}~$\widetilde{\mathcal{B}}_x \! =\! 0.0$~T\end{footnotesize}}
\psfrag{bfdata}[c]{\begin{footnotesize}~$\widetilde{\mathcal{B}}_x \! = \! 1.0$~T\end{footnotesize}}
\psfrag{cfdata}[c]{\begin{footnotesize}~$\widetilde{\mathcal{B}}_x \! = \! 1.5$~T\end{footnotesize}}
\includegraphics[width=80 mm]{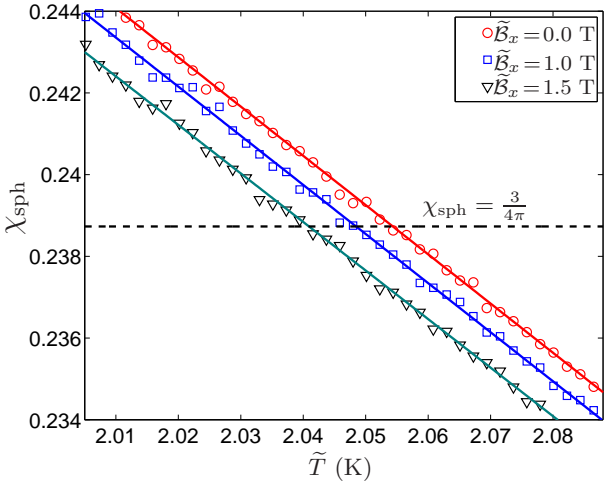}
\caption{Finding $\widetilde{T}_c$ using the perturbative Monte-Carlo
for a sphere of N=295 spins and $J_{\rm ex}=0$,
 by using the reaction field $\chi_{\rm sph}=\frac{3}{4\pi}$ criterion
at criticality	}
\label{JzeroN295}
\end{figure}

   To establish a comparison of the effective perturbative Hamiltonian with previous QMC results \cite{prabuddha}, 
   we first performed 
   Monte-Carlo simulations for a finite size sample with open spherical boundary condition,
    containing $N=295$ spins and with $J_{\rm ex}$ in Eq.~({Risindip})
   set to zero. These conditions are identical to the ones of Ref.~[\onlinecite{prabuddha}].
    As shown in Fig.~\ref{JzeroN295}, similarly to Ref.~[\onlinecite{prabuddha}],  
    we used the reaction field criterion, set by the divergence of $\chi$ 
 when $\chi_{\rm sph}=\frac{3}{4\pi}$
(see Eq.~(\ref{chisphere})),
 to find the effective critical temperature
     $\widetilde{T}_c(\widetilde{B}_x)$ as a function of the effective 
      field $\widetilde{B}_x$, where $\widetilde{T}$ and $\widetilde{B}_x$ are defined 
      in Eqs.~(\ref{effectiveB}) and (\ref{effectiveT}).
      $\chi_{\rm sph}$ is calculated  using
 \begin{eqnarray}
	\chi_{\rm sph}=\frac{1}{k_B\widetilde{T}}\frac{\alpha}{N}\left\langle M_z^2\right\rangle \; ,
\label{chisphere2}
\end{eqnarray}
where the prefactor $\alpha$ is given by 
 \begin{eqnarray}
 \alpha=\frac{4}{a^2c}\left(g_{\rm L}\mu_{\rm B}C_{zz}(0)\right)^2~.
 \end{eqnarray}
 In the perturbative MC method, for determining $\left\langle M_z^2\right\rangle$, we
  used the pseudo-operator defined by Eq.~(\ref{Mz2}).

\begin{figure}[htp]
\center
\psfrag{xl}[c]{$\widetilde{T}$~(K)}
\psfrag{yl}[c]{~~$\widetilde{\mathcal{B}}_x$~(T)}
\includegraphics*[width=90 mm,angle=0]{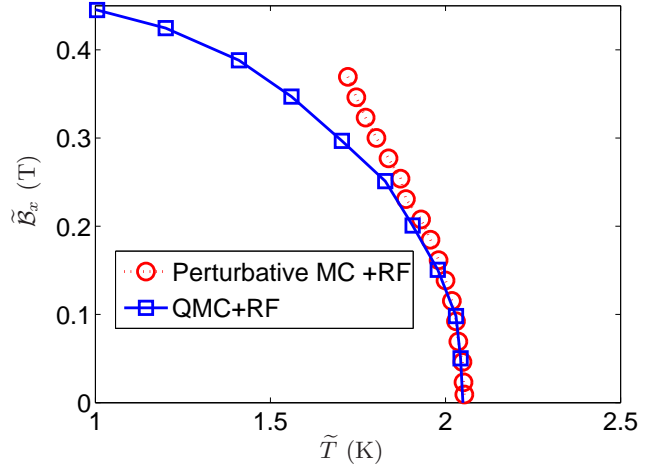}
\caption{\small{Comparing the phase diagram of the perturbative Monte-Carlo
 with Quantum Monte Carlo results \cite{prabuddha} as a function of effective temperature
 and effective magnetic field for a sphere of N=295 spins and $J_{\rm ex}=0$ },
  using the reaction field method of Ref.~[\onlinecite{prabuddha}].}
 \label{Jzero}
\end{figure}

The phase diagram as a function of the effective temperature $\widetilde{T}$ 
and the effective field $\widetilde{B}_x$, using the effective perturbative 
Hamiltonian~(\ref{eff}) and the above cumulant expansion is shown in Fig.~\ref{Jzero}.
It can be seen that at low enough fields close to the classical phase transition, our perturbative
Monte Carlo results, using the same reaction field method as in Ref.~[\onlinecite{prabuddha}], 
closely match the quantum Monte Carlo results from Ref.~[\onlinecite{prabuddha}]. 
Using the reaction field method for $B_x=0$ we get a $\widetilde{T}_c=2.03$~K, where 
$T_c(B_x=0)=\widetilde{T}_c(B_x=0)$ since $\epsilon(B_x=0)=1$.

\subsubsection{Results from Ewald summation method $-$ needle-shaped sample}

The simulations  using the Ewald summation (ES) method were performed
 with simulation boxes of size $L=7,8,9$, with each box containing $N=4\times L^3$ spins.
The dipolar interactions of ions inside the simulation boxes were 
derived via the ES technique and assuming an infinitely long needle-shaped sample~\cite{Tupizin}.
That is, the additional demagnetization term correction from Eq.~(\ref{spher-bound} is not
incorporated into the Ewald representation of the dipolar interactions between
ions $i$ and $j$.
%
 We determined the critical temperature 
 by finding the temperature at which the magnetization Binder ratio~\cite{binder},
\begin{eqnarray*}
 Q=1-\frac{1}{3}\left\langle M_z^4\right\rangle/\left\langle M_z^2\right\rangle ^2	\; , 
\end{eqnarray*}
 for system sizes $L=7,8$,
 and $9$ intersect.
 The intersection point shown in 
 Fig.~\ref{Binder} is
at $T_c=1.92$ K which is the critical temperature.
$\left\langle M_z^4\right\rangle $ and $\left\langle M_z^2\right\rangle$
 are calculated using Eqs.~(\ref{Mz2}) and (\ref{Mz4}) within the perturbative effective Hamiltonian scheme.
As demonstrated in the inset of Fig.~\ref{Binder}, plotting $Q$ as a function of
$L^{1/\nu}(T-T_c)$ shows a good data collapse for system sizes $L=7,8$,
 and $9$, with the mean field exponent $\nu=1/2$.  
 This is consistent with the argument that the upper critical dimension 
 for dipolar interactions is $d=3$. A more rigorous analysis of three dimensional 
 dipolar systems shows logarithmic finite size scaling corrections~\cite{Xu,Larkin}.
 We have not investigated these corrections in this study as it is outside
the scope of this work. As long as $T_c(B_x\ne 0)>0$, the critical behavior should be
controlled by the same classical critical exponents as for $B_x=0$.

\subsubsection{Results from Ewald summation method $-$ spherical sample}

We have repeated the perturbative MC simulations using the ES technique but
 with a slightly different twist to it.
Instead of simulating a long needle-shaped bulk and using the Binder method to obtain 
the critical temperature, 
we simulate a sample with a spherical domain. 
We derived the effective dipolar interactions between
 the spins by using the ES technique for a spherical cavity
The effect of the spherical boundary is 
taken into account by
  incorporating the additional effective
 interaction of Eq.~(\ref{spher-bound})~\cite{vacuum_comment}
   between spins $i$ and $j$.
  Now, one can assume that this sphere is 
embedded in a long-needle-shaped bulk. Therefore, by
recalling the derivation of Eq.~(\ref{chisphere}) from Eq.~(\ref{demagnet}), 
where an effective field
$B_z^{\rm sph}$ is applied to the magnetic moments of the sphere, 
one can determine
 the macroscopic $\chi$
of the bulk, by calculating
$\chi_{\rm sph}$ via ES method for a spherical sample.  
The procedure 
that we use here is similar to the procedure one 
above that employed the reaction field for a finite size system and which led to the
phase diagram in Fig.~\ref{Jzero}.
The difference between the ES technique within a spherical boundary 
and the reaction field method
implemented
\begin{figure}[htp]
\begin{center}
\psfrag{xl}[c]{$\widetilde{T}$~(K)}
\psfrag{yl}[c]{~~\large{$Q$}}
\psfrag{x2}[c]{$\widetilde{T}$~(K)}
\psfrag{y2}[c]{~~\large{$\chi_{\rm sph}$}}
\psfrag{L2}[c]{$L^{1/\nu}(\widetilde{T}-\widetilde{T}_c)$}
\psfrag{Q}[c]{~~\large{$Q$}}
\psfrag{c2}[c]{~~$\chi_{\rm sph}=\frac{3}{4\pi}$}
\includegraphics*[width=8 cm,angle=0]{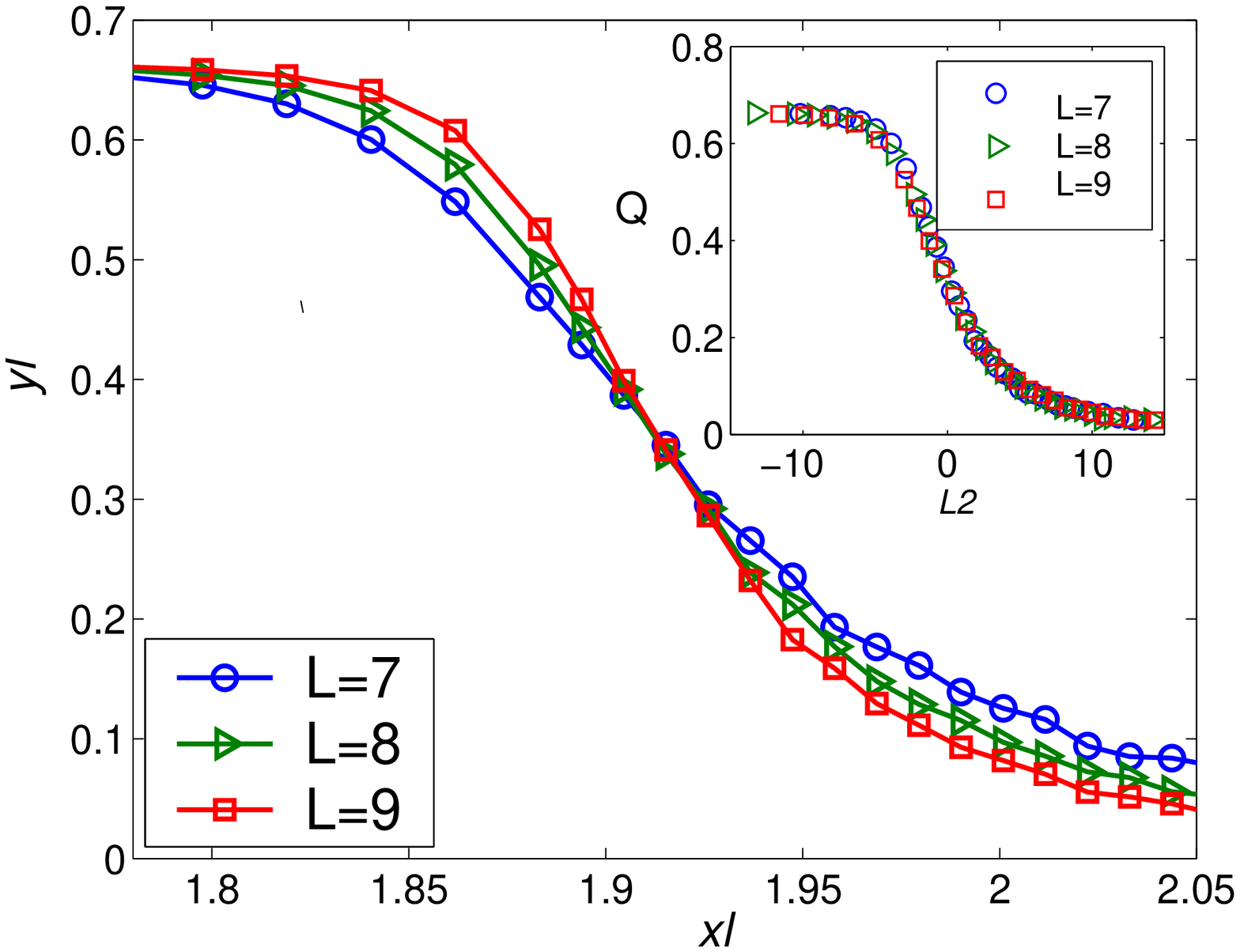}
\includegraphics*[width=8 cm,angle=0]{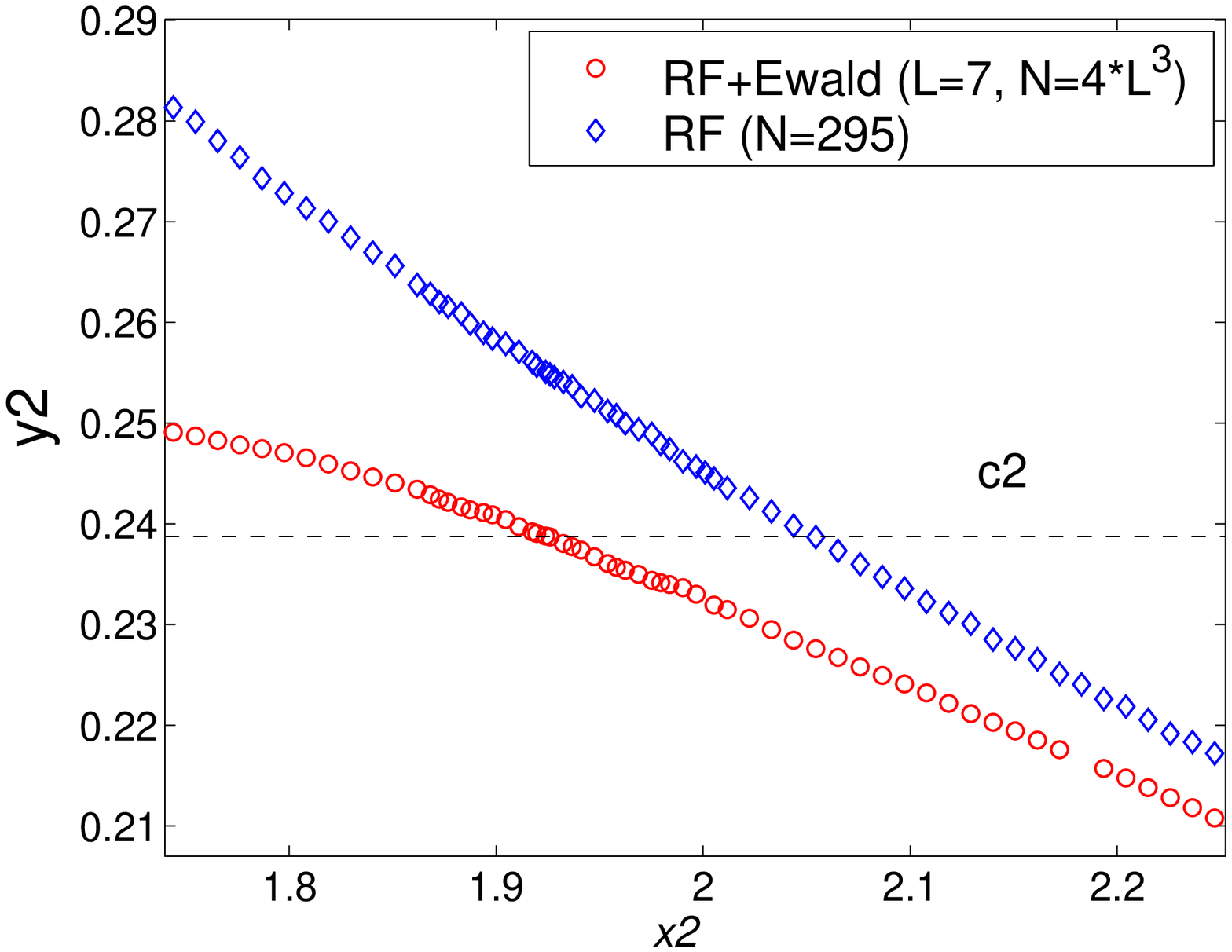}
\caption{\small{(a) The Binder ratio crossing for $L=7,8,9$ system sizes,
 performing MC and using ES technique for a long needle-shaped sample, 
 with $B_x=0$, $J_{\rm ex}=0$. $\widetilde{T}=T$ for $B_x=0$. The inset shows that the Binder ratios collapse for the mean field
exponent $\nu=1/2$ to a very good degree.  
(b) $\chi_{\rm sph}$ calculated by performing MC simulation, using Eq.~(\ref{chisphere2}). 
The diamonds are for a finite
size sphere using the reaction field scheme similar as in  Ref.~[\onlinecite{prabuddha}]
(i.e. same results as shown in Fig.~\ref{JzeroN295} for the $\widetilde{\mathcal{B}}_x =0$ data).
For the circles, we have obtained the interaction between the ions by the ES technique for $L=7$
system size and incorporating
the spherical boundary effect via the demagnetization term of
 Eq.~(\ref{spher-bound}) and using $B_x=0$ and $J_{\rm ex}=0$, 
with again $\widetilde{T}=T$ for $B_x=0$. As one can see the $\widetilde{T}_c \approx 1.92$ K
 obtained here 
agrees with 
the $\widetilde{T}_c$ obtained using the Binder ratio crossing.}}
 \label{Binder}
 \end{center}
\end{figure}
in Ref.~[\onlinecite{prabuddha}] is that instead of using an open spherical boundary condition,
and considering only 
bare dipolar interaction between a finite number of spins within a cutoff sphere,  
a simulation box with periodic boundary condition is considered.
The effective dipolar interactions of ions inside the simulation box is 
derived via the Ewald summation technique. In this approach a spherical boundary 
is considered for the whole simulation box and all the replicated images
of the real box. In this case,
each effective pairwise 
dipolar interactions described by the ES representation
has added to it the extra interaction term given by Eq.~\ref{spher-bound}.
Once again, the origin of this additional interaction is  the demagnetization field, 
due to the polarization of the magnetic moments on the spherical boundary. 
In this approach, the system behaves much more like a
macroscopic sphere  compared to 
the one above that used the reaction field method.
 It is further assumed that this macroscopic sphere is embedded inside a macroscopic 
 long macroscopic needle-shaped domain. Therefore, by employing the perturbative Monte Carlo method
  and using Eq.~(\ref{chisphere2}), we calculate $\chi_{\rm sph}$ to obtain 
 the critical temperature. Based on Eq.~({\ref{chisphere}), the critical temperature is calculated by 
 finding where the
 $\chi_{\rm sph}=\frac{3}{4\pi}$ 
criticality criterion is satisfied. 
As shown in Fig.~\ref{Binder}b, for a simulation box of $L=7$, we obtain $T_c=1.92$~K 
for a zero transverse field and $J_{\rm ex}=0$,
very close to the $T_c$ previously derived using ES technique for a long needle-shaped sample and
shown in Fig.~\ref{Binder}a. 
Thus, the two approaches using ES technique lead to similar results.
We believe that the difference between the classical $T_c$ obtained via 
ES technique and the $T_c(B_x=0)$ obtained using the reaction field method ~\cite{prabuddha} is because,
in the reaction field method, the 
number of spins inside the cut-off sphere, which is embedded in the needle-shaped domain,
is of too limited size. 
By implementing  Eq.~(\ref{chisphere}) 
in the reaction field method,
 the effect 
of the spins on the spherical boundary for a limited size is in essence
incorporated
in a mean field manner in the simulation. For a limited size boundary,
thermal fluctuations on the boundary are
underestimated,
 hence resulting in an overestimated $T_c$.
  This overestimation of $T_c$,
 which decreases by increasing the size of the spherical boundary, 
  is expected to vanish in the thermodynamic limit $L\rightarrow\infty$.  
\subsection{Nearest-Neighbor Exchange Interactions}
The zero transverse field critical temperature of 1.92 K obtained above lies quite
far above
the experimental critical temperature of $1.53$~K. 
As suggested by Chakraborty {\it et al.}, 
it is reasonable to assume that the discrepancy may be related to a nearest-neighbor
 Heisenberg antiferromagnetic exchange interaction. Indeed, in the related LiTbF$_4$ material, 
 it has long been known that a $J_{\rm ex}$ coupling exists~\cite{LiTb}.
 There has been no direct determination for the magnitude of 
 this nearest-neighbor exchange in LiHoF$_4$. However, 
 there have been indirect estimations, considering $J_{\rm ex}$ as a free parameter, such that
  the specific heat~\cite{Mennenga} and susceptibility~\cite{Beauv} calculations based on mean field theory
  fit to the equivalent experimental measurements. Another procedure to determine $J_{\rm ex}$,
   would be to fit theoretical calculation with neutron scattering data, similar to the procedure
  followed for LiTbF$_4$~\cite{LiTb}. Recently, R\o nnow \textit{et. al}~\cite{Ronnow} have
   performed inelastic neutron scattering
  measurements on LiHoF$_4$. Considering $J_{\rm ex}$ as a free parameter, they used the so called
  \textit{effective-medium theory} to modify the mean field random phase approximation parameters.
  They estimated $J_{\rm ex}$ such that a best fit with the experimental phase diagram is
   obtained.
  For example, although for $J_{\rm ex}$=$1.16$~mK there is good agreement with experiment
  when $2.0 < B_x < 4.0$~Tesla, as is common in mean field theory calculations,  
  the critical temperature is overestimated (by 14 percent) compared with the experimental critical temperature
  at zero applied field $B_x=0$.  
  
  In our work here, we use Monte Carlo techniques
   and consider the exchange interaction as a free parameter.
   We can estimate the $J_{\rm ex}$ strength by adjusting its value  
   such that the experimental $T_c$ is reproduced, as was done in Ref.~[\onlinecite{prabuddha}]. 
\begin{figure}[htp]
\center
\psfrag{T}[c]{$\widetilde{T}$~(K)}
\psfrag{T2}[c]{$\widetilde{T}$~(K)}
\psfrag{c}[c]{$\chi_{\rm sph}=\frac{3}{4\pi}$}
\psfrag{chi}[c]{~~\large{$\chi_{\rm sph}$}}
\psfrag{Q}[c]{~~\large{$Q$}}
\includegraphics*[width=\columnwidth,angle=0]{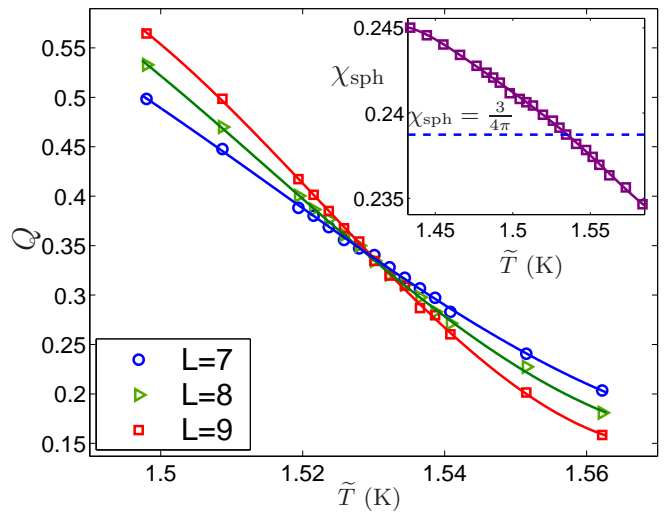}
\caption{\small{The Binder ratio crossing for $L=7,8,9$ system sizes, performing MC 
and using ES technique for a cylindrical boundary with $B_x=0$. $J_{\rm ex}=6.07$~mK is set such that
the critical temperature $T_c\approx 1.53$~K is obtained. $\widetilde{T}=T$ for $B_x=0$.
In the inset $\chi_{\rm sph}$ is calculated by performing Monte Carlo simulations, 
using Eq.~(\ref{chisphere2}). The interaction between the ions is obtained by the ES technique 
for $L=7$ system size and using a spherical boundary condition for $B_x=0$. The same
 $J_{\rm ex}=3.91$~mK used and a similar $T_c\approx1.53$~K is obtained.
  $\widetilde{T}=T$ for $B_x=0$.}
 \label{J557}}
\end{figure}
Using the reaction field method performed for finite spheres in Ref.~[\onlinecite{prabuddha}], 
for $N=295$ spins, 
$J_{\rm ex}=6.07$~mK was obtained. 
As a check, we repeated our
Monte Carlo simulations, also using the reaction field method 
for the same number of spins, and fitted
$J_{\rm ex}$ such that the experimental 
 zero-field critical temperature $T_c=1.53$~K is reproduced. 
 We obtained the same  $J_{\rm ex}=6.07$~mK as in Ref.~[\onlinecite{prabuddha}].
  It should be noted that, as reported in Ref.~[\onlinecite{prabuddha}],
  one does not obtain a unique $J_{\rm ex}$ value when performing simulations for different 
  sphere sizes.
The $J_{\rm ex}$ value strongly depends on the number of spins considered. 
In Ref.~[\onlinecite{prabuddha}], for the largest system size considered (N=3491), 
a $J_{\rm ex}=5.25$~mK was required to obtain a Monte Carlo estimate of $T_c$ of 1.53~K.
\begin{figure}[htp]
\center
\psfrag{x1}[c]{$\widetilde{T}$~(K)}
\psfrag{y1}[c]{~~\large{$Q$}}
\psfrag{x2}[c]{$\widetilde{T}$~(K)}
\psfrag{y2}[c]{~~\large{$Q$}}
\includegraphics*[width=7.5 cm,angle=0]{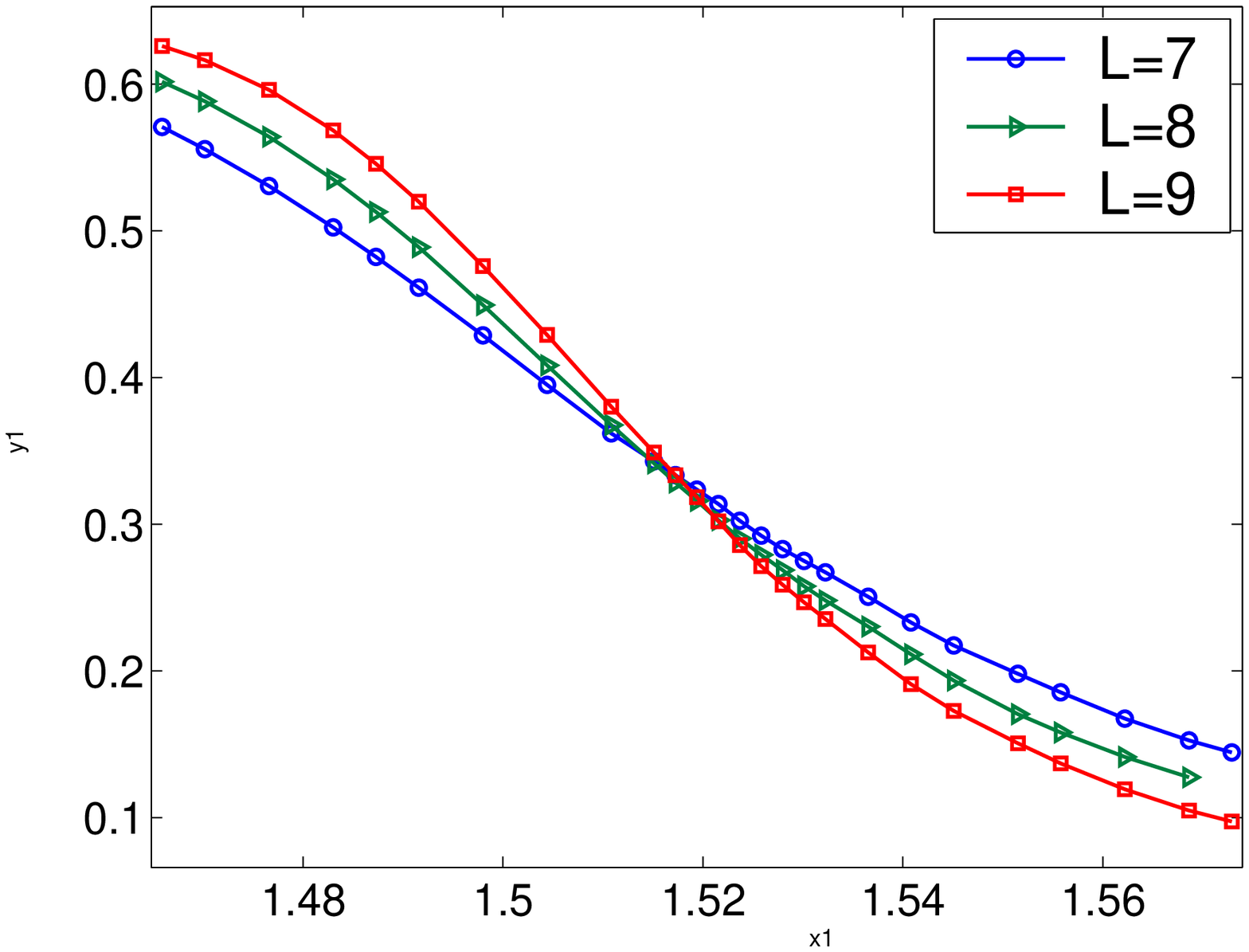}
\includegraphics*[width=7.5 cm,angle=0]{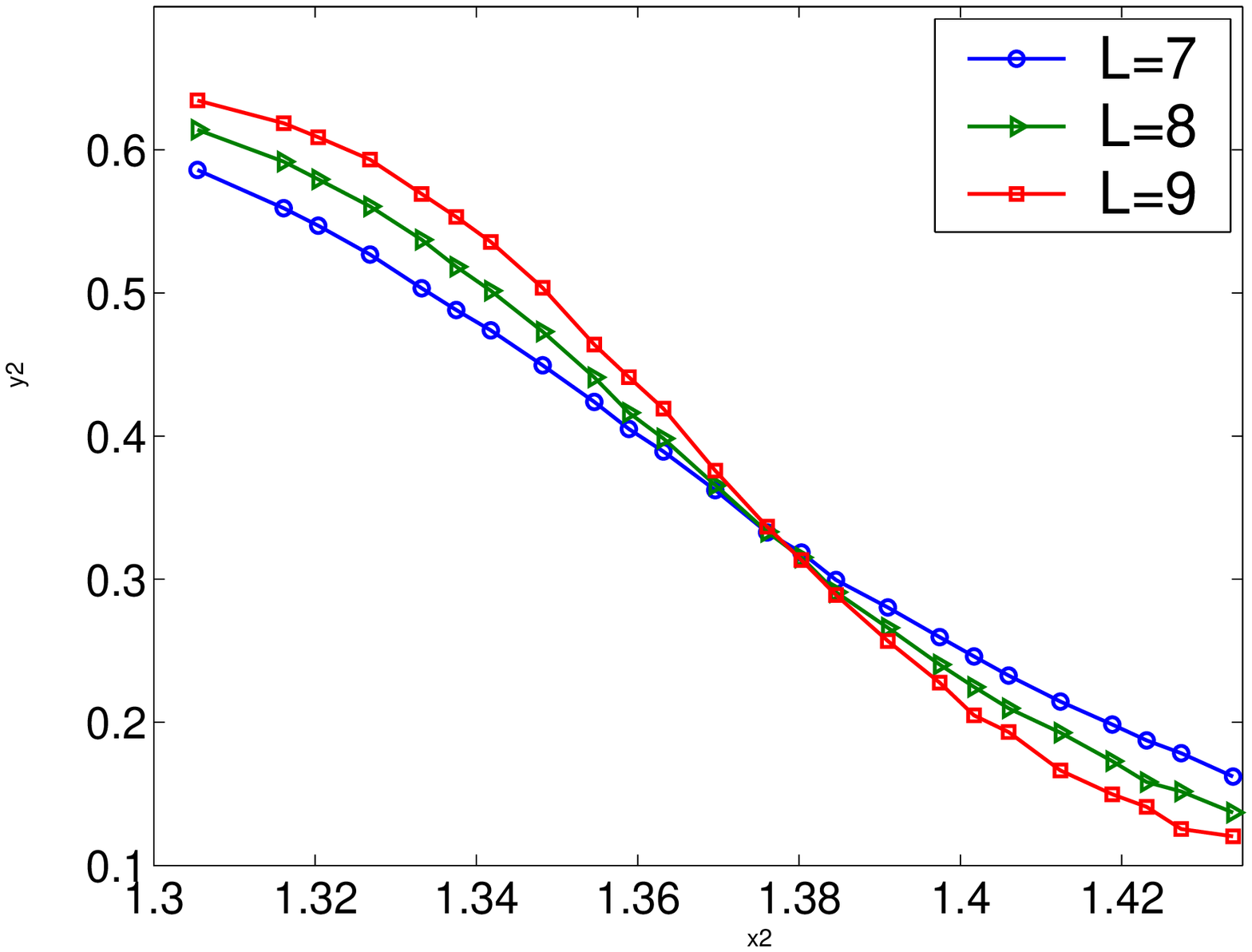}
\caption{\small{The Binder cumulant crossing for $L=7,8,9$ system sizes,
 performing perturbative MC and using
 ES technique for a long needle-shaped sample with $J_{\rm ex}=3.91$~mK. 
In (a) we have $\widetilde{B}_x=0.05$~T and in (b) we have $\widetilde{B}_x=0.15$~T.}}
 \label{BinderJ557}
\end{figure}
There are two sources of errors that are affecting the value of the estimated $J_{\rm ex}$ 
obtained by the reaction field method of Ref.~[\onlinecite{prabuddha}].
 Firstly, for a given number of spins, when Monte Carlo simulations are performed
   to calculate $T_c$, the reaction field method estimates a higher value for $T_c$ compared to the ES  method.
   The sources of these errors are finite size effects and the underestimation of thermal fluctuations
at the boundary,
as we now explain. 
   To push down the value of $T_c$ obtained for $J_{\rm ex}=0$ such that it matches the
 experimental value for $T_c$, an antiferromagnetic $J_{\rm ex}$
is required. For $J_{\rm ex}=0$, the reaction field method generates a higher $T_c$ compared to the ES method.
Therefore,
 in order to push down the $T_c$ obtained from Monte Carlo simulation to match the
 experimental value for $T_c$, a larger value for the antiferromagnetic $J_{\rm ex}$
is required than the one required when using the ES method.
Secondly, there is another source of error affecting the value of the estimated $J_{\rm ex}$ 
obtained by the reaction field method. It comes from
the number of 
 surface bonds,
which depends on the radius of the chosen cut-off sphere. For ions close to the 
surface, some of the nearest-neighbors 
fall inside the spherical boundary while some remain outside. Because of the missing 
 number of exchange interactions on the boundary, the overall exchange estimated is 
forced to be larger than the 
actual value. When the ES technique is used in conjunction with periodic boundary 
conditions, this boundary effect problem no longer exists, making the ES technique 
a more reliable tool for estimating 
$J_{\rm ex}$~\cite{spinice,Melko}.
To estimate $J_{\rm ex}$ using our Monte Carlo simulations,
we used the Binder ratio crossing method and employed
both the ES technique for a long needle-shaped sample and
the ES technique for a macroscopic sphere embedded in a long needle-shaped sample. 
For the latter case, the interactions of
 Eq.~(\ref{spher-bound}),
originating
 from the magnetic polarizations of the magnetic moments on the spherical boundary were considered as well.   
The two $J_{\rm ex}$ values so determined
are the same, which 
is approximately  $J_{\rm ex}=3.91$~mK, as illustrated in Fig.~\ref{J557}. 
Note that this value of  $J_{\rm ex}=3.91$~mK is consistent with
the one recently determined in Ref.~[\onlinecite{Biltmo}].
The definition of the
exchange constant of 0.12 K in Ref.~[\onlinecite{Biltmo}] for 
Ising spins corresponds to 
$J_{\rm ex}{C_{zz}(B_x=0)}^2$ in our case.
Using
$J_{\rm ex}=3.91$~mK and
 $C_{zz}(B_x=0)=5.51$
from Fig.~\ref{evolution},
we have $J_{\rm ex}{C_{zz}(B_x=0)}^2 \approx 0.119$~K, in
excellent agreement with Ref.~[\onlinecite{Biltmo}].

\subsection{Transverse Field vs Temperature Phase Diagram}

Having determined a seemingly consistent value for $J_{\rm ex}$, we are now ready to perform 
Monte Carlo simulation for 
small transverse magnetic fields $B_x$. The effect of quantum perturbations are 
incorporated through the effective 
Hamiltonian Eq.~(\ref{eff}), which is derived from the $B_x-$rescaled
Hamiltonian Eq.~(\ref{Risingdip}).
To obtain the real temperature $T$ and external transverse magnetic field $B_x$ from the
 effective values $\widetilde{T}$ and $\widetilde{B}_x$
used in the simulations
 we employ relations
Eqs.~(\ref{effectiveB}) and (\ref{effectiveT}). To illustrate the procedure, we show the 
crossing of the Binder ratio $Q$ for $\widetilde{B}_x=0.05$~T 
and $\widetilde{B}_x=0.15$~T in Fig.~\ref{BinderJ557}.
 \begin{figure*}[t]
 \center
\psfrag{T}[c]{$T$~(K)}
\psfrag{Bx}[c]{$B_x$~(T)}
\psfrag{Exp}[c]{\begin{footnotesize}~~~~~~~~~~~Experiment~(~Ref.~[\onlinecite{Bitko}]~)\end{footnotesize}}
\psfrag{Girvin}[b]{\begin{footnotesize}QMC+RF, $N \! = \! 295$, $J_{\rm ex} \! = 
\! 6.07$~mK\end{footnotesize}}
\psfrag{PMC}[c]{\begin{footnotesize}~PMC+RF, $N \! = \! 295$, $J_{\rm ex} \! = \! 6.07$~mK\end{footnotesize}}
\psfrag{RFew}[c]{\begin{footnotesize}PMC+ES (spherical), $J_{\rm ex} \! = \! 3.91$~mK\end{footnotesize}}
\psfrag{Bind}[c]{\begin{footnotesize}~PMC+ES (needle-shaped), $J_{\rm ex}\! = \! 3.91$~mK\end{footnotesize}}
\includegraphics [width=120 mm,angle=0,keepaspectratio]{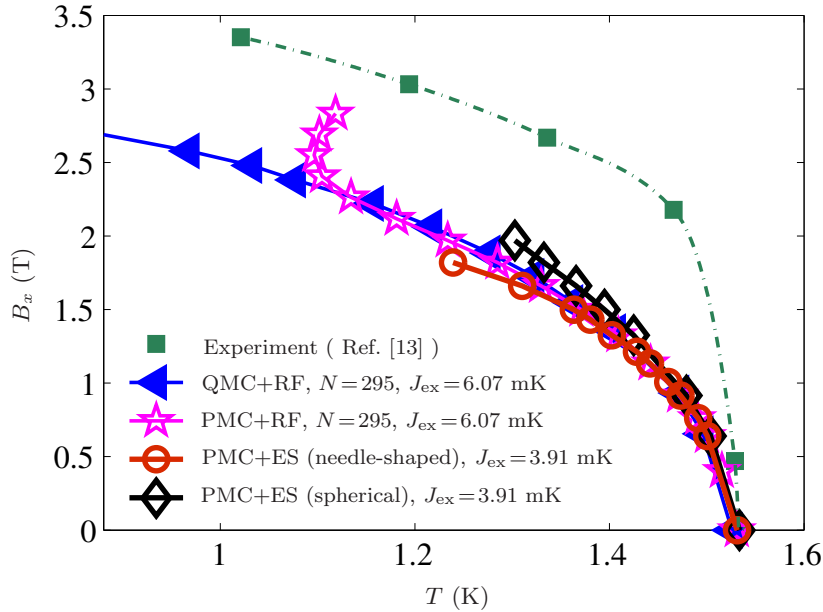}
\caption{\small{The phase diagram of the critical transverse field as a
 function of temperature for LiHoF$_4$.
 The closed boxes are the experimental phase digram \cite{Bitko}. 
 The closed triangles are the phase diagram obtained by QMC \cite{prabuddha}
  using the RF method for a finite sphere with $N=295$ spins.
 The open stars are the result from perturbative Monte Carlo (PMC) using the same 
 reaction field (RF) method used in Ref.~[\onlinecite{prabuddha}] for a sphere with $N=295$ spins.
 Quite importantly, as discussed in the text, the reaction field method leads to a considerable
 overestimate of $J_{\rm ex}$.
 The open circles are obtained, using the perturbative Monte Carlo 
 in a needle-shaped domain using ES method.   
 The open diamonds are obtained, using perturbative 
 Monte Carlo in a bulk sphere embedded in a needle-shaped domain, 
 using ES method and the spherical boundary effect of Eq.~(\ref{spher-bound}). }}
 \label{everything}
\end{figure*} 

Interestingly, using each of the numerical methods discussed above to obtain the phase diagram,
 it seems that for small $B_x$ the final phase diagrams demonstrating the critical transverse field as a
  function of temperature
 are affected very little in respect to which specific 
technique is used. Figure \ref{everything}
 shows the phase diagrams, using the perturbative Monte Carlo method
 implementing the reaction-field method and the Ewald summation technique,
 compared with QMC \cite{prabuddha} results and experiment \cite{Bitko}.
 We use Eq.~(\ref{effectiveB}) and Eq.~(\ref{effectiveT}) to obtain 
 the real physical transverse magnetic field, $B_x$ and temperature $T$ from
 $\widetilde{T}$ and $\widetilde{B}_x$.
 As one can see, all the phase diagrams obtained from the effective perturbative
 method show a good agreement with the  
 quantum Monte Carlo result of Chakraborty {\it et al.}~\cite{prabuddha},
  for small transverse fields
  up to a ``real'' physical transverse magnetic field $B_x \approx 1.5$~Tesla, where we 
  presume 
the lowest order cumulant formulation of the effective classical Hamiltonian
model breaks down. This is the main result of this work. 

In conclusion,  we confirm the results of Ref.~[\onlinecite{prabuddha}] but, perhaps unfortunately, 
  we fail to explain the discrepancy between numerical and experimental results. We are thus led to ponder 
  on theoretical reasons that may explain this discrepancy. We explore one such possibility in the next subsection
  and which is also the one that was put forward in Ref.~[\onlinecite{prabuddha}].

 \subsection{Other Crystal Field Parameters }

 As reported in Ref.~[\onlinecite{prabuddha}],
we find
 that the numerical phase diagrams show a discrepancy with the experimental phase 
 diagram, even at asymptotically small transverse fields.
 Indeed, this was one of the main motivations for the present work.
 As can be seen in Fig.~\ref{everything}, our efforts
in considering (i) a different
 Monte Carlo scheme and (ii) other ways to handle the long-range dipole-dipole interactions 
have not allowed us to resolve the
  discrepancy between the results from numerical simulations of Ref.~[\onlinecite{prabuddha}]
  and the experimental phase diagram of Ref.~[\onlinecite{Bitko}].
  Chakraborty {\it et al.}~\cite{prabuddha} suggested that this discrepancy may be related 
  to uncertainties in the crystal field parameters.
We now briefly explore this possibility.

   As discussed in Appendix~\ref{A}, 
  crystal field parameters are usually obtained such that theoretical calculations match
  with experimental data from electron paramagnetic resonance (EPR)~\cite{Malkin}, 
  inelastic neutron scattering (INS)~\cite{Ronnow} or
   susceptibility measurements~\cite{Hansen}. 
Recalling the discussion that led to the derivation of the effective spin-$1/2$ description of LiHoF$_4$ 
in Eq.~(\ref{isingdip}), one realizes that the parameters $C_{zz}(B_x)$ and $\Delta(B_x)$ are implicit functions 
of the crystal field level
energies and crystal field level
wave functions. As a result, the mapping of the problem to a
spin-$1/2$ model depends on the chosen values of the $B_n^{\alpha}$ crystal field parameters (See Appendix \ref{A})
entering in the description of the crystal field Hamiltonian $V_c$. This state of affairs is rendered
particularly
 important, since, unfortunately, there appears to be some ambiguity in the literature about the empirical values
  of the $B_n^{\alpha}$ parameters.
 All the numerical results that we obtained in the previous sections
  are based on the set of recent
  crystal field parameters obtained reported in Refs.~[\onlinecite{Ronnow}], 
which were also used in the stochastic series expansion quantum Monte Carlo
of Ref.~[\onlinecite{prabuddha}], and which
  were determined by fitting theoretically determined crystal field levels 
  with those resolved in inelastic neutron scattering data.
  Recently, new electron paramagnetic resonance (EPR) spectroscopy experiments have been 
  performed, in which the crystal field
  parameters were determined~\cite{Malkin}. 
 Based on the EPR data reported in Ref.~[\onlinecite{Malkin}], spectral
 parameters were refined in order to fit the observed dependencies of the resonance frequencies on the
  external magnetic field, giving a new set of crystal field parameters and an 
  effective Land\'e g-factor $g_{\rm L}$ reduced from its pure $^5I_8$  
$g_{\rm L}=5/4$ value 
  down to an effective 
$g_{\rm L}^{\rm eff}=1.21$.
   Using this new set of crystal-field parameters, 
   we obtain a
  different renormalization factor $\epsilon(B_x)$ (Eq.~(\ref{epsilon})) and effective
   transverse field $\widetilde{\mathcal{B}}_x$ (Eq.~(\ref{effectiveB})) and as a result, 
    different $C_{zz}(B_x)$ and $\Delta(B_x)$.
   \begin{figure}[h]
   \center
   \psfrag{B}[c]{$B_x$~(T)}
   \psfrag{B2}[c]{$B_x$~(T)}
   \psfrag{C}[c]{$C_{zz}$}
   \psfrag{D}[c]{$\Delta$~(K)}
\includegraphics*[width=\columnwidth,angle=0]{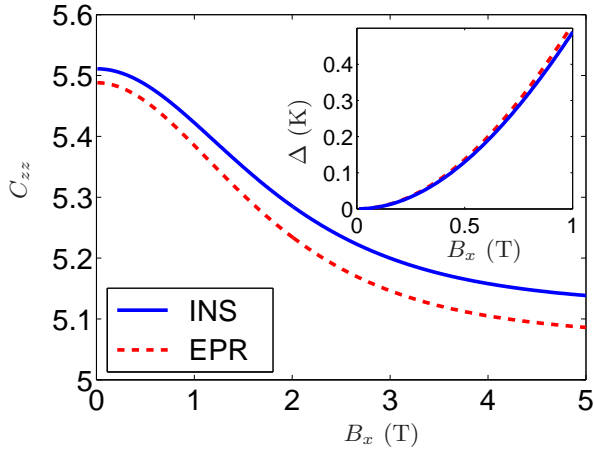}
\caption{\small{Comparing $C_{zz}$ and $\Delta$ as a function of $B_x$, 
calculated using two different crystal field parameters
 (CFP). The solid line is obtained, using Refs.~[\onlinecite{prabuddha,Ronnow}] CFP based on inelastic 
 neutron scattering (INS) experiment.
 The dashed lines is obtained, using Ref.~[\onlinecite{Malkin}] based on electron 
 paramagnetic resonance (EPR) experiment.}}
 \label{malkin}
\end{figure} 
\begin{figure}[h]
 \center
\psfrag{T}[c]{$T$~(K)}
\psfrag{Bx}[c]{~~~$B_x$~(T)}
\psfrag{data1}[c]{~~~~\begin{footnotesize}Experiment (~Ref.~[\onlinecite{Bitko}]~)\end{footnotesize}}
\psfrag{Girvin}[b]{\begin{footnotesize} MC+RF, $N=295$, $J_{\rm ex}=6.07$~mK\end{footnotesize}}
\psfrag{PMC}[c]{\begin{footnotesize}PMC+RF, $N=295$, $J_{\rm ex}=4.38$~mk\end{footnotesize}}
\includegraphics [width=9 cm,angle=0,keepaspectratio]{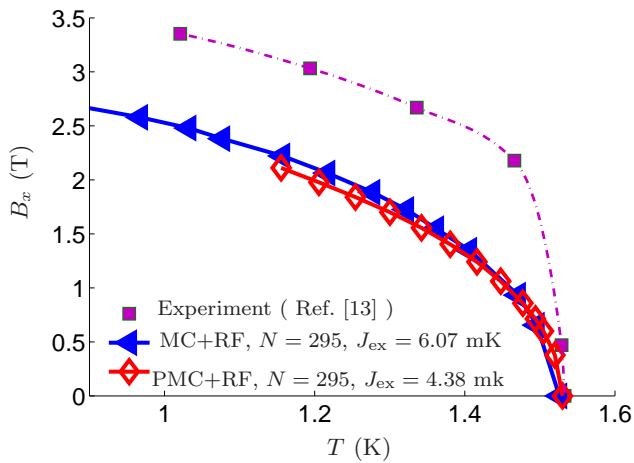}
\caption{\small{Comparing the phase diagrams of the critical transverse
 field as a function of temperature for LiHoF$_4$, based on two different set of 
 crystal field parameters. The closed triangles are the QMC results of Ref.~[\onlinecite{prabuddha}],
  using the RF method for a finite sphere with $N=295$, based on the CFP of Ref.~[\onlinecite{Ronnow}].  
The open diamonds are obtained from our perturbative Monte-Carlo, using the same RF method used in 
Ref.~[\onlinecite{prabuddha}] for a sphere with $N=295$ spins, 
based on the CFP reported in Ref.~[\onlinecite{Malkin}].}}
 \label{everythingMal}
\end{figure} 
  One of the consequences of obtaining a different $C_{zz}$, 
  with the new CFP is that referring, to Eq.~(\ref{isingdip}), 
  a different $B_x=0$, $T_c$ is obtained. Having 
determined a different $T_c$ via this new set of CFP, 
  the value of  $J_{\rm ex}$ 
required to match
 the experimental $T_c=1.53$~K
  is therefore different.
   In order to scrutinize ``only" the effect of using a 
new set of CFP and to compare the phase diagram 
obtained using this new set of parameters with the results of Ref.~[\onlinecite{prabuddha}] in a rather simple way, 
  we repeated the perturbative Monte Carlo simulations,
 using the same reaction field method used above and done
 in Ref.~[\onlinecite{prabuddha}] for
a finite size sphere of $N=295$ spins 
 and a newly determined $J_{\rm ex}=4.38$~mK.  
At the end, after in essence repeating all the work discussed in Section~\ref{RF-method-results}, 
 a new phase diagram is derived. This phase diagram is plotted in Fig.~\ref{everythingMal}. 
 As it can be seen, this \textit{new} phase diagram
  is consistent with the previous theoretical work, (\textit{e.g.} Ref.~[\onlinecite{prabuddha}] and
 Fig.~\ref{everything}). Interestingly it therefore does not appear at this time that the 
 different crystal field Hamiltonians available for
 LiHoF$_4$~\cite{prabuddha,Ronnow,Malkin,Hansen} are able to explain the significant discrepancy between the $B_x-T$
 phase diagram obtained by simulations compared to experimental results of Ref.~[\onlinecite{Bitko}].
 Finally, it should be emphasized that there is no difference in the results for this new set of CFP 
  provided $J_{\rm ex}$ is adjusted as well.
 On the other hand, different CFP
 lead to a systematically  different $T_c$ if $J_{\rm ex}$ is not adjusted.   

\section{Conclusion}

With a perturbative Hamiltonian derived from a low energy effective spin-$1/2$ 
description of 
LiHoF$_4$, we have re-investigated the $B_x-T$ phase diagram with an
 independent approach for small
 $B_x/T_c$ where quantum fluctuations are weak. 
 The method we used to incorporate perturbatively weak quantum fluctuations within a
 semiclassical Hamiltonian, because of its simple numerically tractable form, 
 allows one to directly address possible factors behind the discrepancy between results from experiments
 and from classical Monte Carlo simulations in the vicinity of $T_c$. This method can be easily generalized to
  more complicated quantum magnetic Ising models, where the Ising-like term is the dominant term 
  and the other non commuting terms are considered as weak perturbations. 
  In particular, if one is interested in studying numerically the effect of nonzero $B_x$ in the diluted regime of LiHo$_x$Y$_{1-x}$F$_4$,
   this perturbative method should be directly applicable 
  by performing Monte Carlo simulations of the appropriate low energy Hamiltonian~\cite{Tabei-PRL, Tabei-Vernay}. 
 
 To perform semi-classical Monte Carlo simulations that handle
 the magnetostatic long-range dipole-dipole interactions properly, we applied the 
 Ewald summation technique for two different geometries.
 In order to determine
$T_c$, 
we used the Binder magnetization ratio crossing for a 
 long needle-shape sample, and we used the $\chi_{\rm sph}=\frac{3}{4\pi}$ criterion
  for a spherical sample
 embedded inside a long
 needle-shaped domain. 
We obtained the same $T_c$ for both cases and, consequently, 
determined the same value for $J_{\rm ex}$. 
 The values of the $T_c$ and $J_{\rm ex}$ that we calculated
  are somewhat
different from the $T_c$ and  $J_{\rm ex}$ values found
 in Ref.~[\onlinecite{prabuddha}].
  This difference originates from using open boundary conditions 
 and a finite spherical cutoff in Ref.~[\onlinecite{prabuddha}], 
 which underestimates the
thermal fluctuations at the boundary.
 We found that although we used a different method and found a different $J_{\rm ex}$, 
  the final $B_x-T$ phase diagram obtained here is the same in the low $B_x/T_c$ limit
 as in the previous results~\cite{prabuddha}. As a result, we tentatively conclude
that the discrepancy between the theoretical and experimental results is {\it not} of
computational origin. 
To explore a possible explanation for the discrepancy, we considered a different set of
crystal
field parameters.
 
A consideration of different
crystal field parameters (CFP), which  lead to 
a different estimate for $J_{\rm ex}$  does not, however,
 at the end produce a dramatically different $T_c$ vs $B_x$ phase diagram. 
 This preliminary result that only considers one set of alternative CFP goes against the suggestion
  of Ref.~[\onlinecite{prabuddha}] that a
 possible origin of the
  discrepancy might be due to the ambiguity in CFP.
  It is perhaps surprising that the consideration of a rather different set of CFP
    compared to those used in Ref.~[\onlinecite{prabuddha}] affects the phase diagram so little once
     $J_{\rm ex}$ has been re-adjusted
to match the experimental $T_c(B_x=0)=1.53$~K value.
 Therefore the origin of the discrepancy between numerics and experiment remains 
fully unexplained.

 The method we obtained in the present work could
be used to carry on further 
investigation of 
the cause of the discrepancy. Without this tool, it would have been somewhat 
less straightforward to have investigated the relevance of the various
factors that we investigated in this paper.
The disagreement  with the 
experimental phase diagram of Ref.~[\onlinecite{Bitko}], would suggest that it may be
 worthwhile 
to revisit the experimental determination of the $B_x$ vs $T_c$ phase diagram.
On the other hand, in both the work presented here and in that of Ref.~[\onlinecite{prabuddha}],
a very simple spin Hamiltonian was considered. Specifically, only long-range
 magnetostatic dipole-dipole 
and isotropic (Heisenberg) nearest-neighbor exchange interactions were considered.
The faster decreasing $T_c(B_x)$, compared to the experimental
 case indicates that perhaps there are effects at play in the
 real material that weaken quantum fluctuations for small $B_x$.

In other words, there may be other couplings in the effective theory in addition to those
in the simplest transverse field Ising model (TFIM) of Eq.~(\ref{HOneHalf}).
As illustrated in Fig.~\ref{ratio}, the terms that we ignored when passing 
from Eq.~(\ref{full2D}) to Eq.~(\ref{isingdip}) seem too small to be able to resolve this issue.
It might be necessary to consider the possibility that not completely negligible  
anisotropic exchange, higher order multipolar exchange 
interaction, or magneto-elastic couplings may be at play in LiHoF$_4$. 

Finally, we note that it would be interesting if one 
could study other magnetic materials similar to the LiHoF$_4$ compound and that could 
provide 
another realization of a TFIM.
   Recently, a mean-field theory calculation has concluded that
 Ho(OH)$_{3}$, which is an insulating hexagonal dipolar Ising ferromagnet, 
 is very well described by a TFIM when a magnetic field $B_x$ is applied perpendicular to the 
Ising spin direction~\cite{Pawel}. This material constitutes  
a close analogue of LiHoF$_4$ and, when diamagnetically diluted with Y$^{3+}$,
may potentially be an analogue of LiHo$_x$Y$_{1-x}$F$_4$. 
The existence of another experimental
 candidate for the study of the TFIM with long-range dipolar interaction presents the opportunity 
 to re-investigate the puzzling properties of pure and diluted LiHoF$_4$ in a new material,
 shedding light on the physics of dipolar Ising systems in both zero and nonzero applied transverse field.
 The method we have employed in this work is a suitable tool to study these new proposed quantum magnetic Ising materials
 beyond mean field theory and provides a tool to make comparison with future experiments performed on these proposed TFIM materials. 
 
To conclude, we hope that the work presented here stimulates further theoretical and 
experimental studies of LiHoF$_4$ in the regime of small transverse field
 $B_x$ where the classical paramagnetic to ferromagnetic transition 
 is only perturbatively affected by $B_x$. 
\section{Acknowledgments}
We thank
P. Chakraborty, J.-Y. Fortin, S. Girvin,
P. Henelius,
 R. Hill, B. Malkin, P. McClarty, J. Quilliam, P. Stasiak, and  F. Vernay for 
 useful discussions. 
Support for this work was provided by the NSERC
of Canada and the Canada Research Chair Program (Tier I, M.G),
the Canada Foundation for Innovation,
the Ontario Innovation Trust, the Canadian 
Institute for Advanced Research, and NSC of Taiwan.
\appendix
\section{}\label{A}
In this Appendix we briefly discuss how the crystal field Hamiltonian of LiHoF$_4$ is written
in terms of angular momentum operators and crystal field parameters. 

In the point charge approximation description of the crystal field, we assume that the ions interacting with
Ho$^{3+}$ electrostatically are close to point charges. The
potential at $\mathbf{r}$ is simply the sum of point charge coulomb interaction potential 
\begin{equation}
V(\mathbf{r}) = \sum_i \frac{q_i}{|\mathbf{R}_i - \mathbf{r}|}~,
\end{equation} 
where $\mathbf{R}_i$ is the position and the total electric charge of the $i$'th ion. 
 $V(\mathbf{r})$ can be expanded as
\begin{equation} V(r,\theta,\phi) = \sum_{n=0}^\infty \sum_\alpha
r^n \gamma_{n\alpha} Z_{n\alpha}(\theta,\phi)~, 
\end{equation} 
where
\begin{equation} 
\gamma_{n\alpha} = \sum_i \frac{4\pi q}{(2n+1)}\frac{Z_{n\alpha}(\theta_i,\phi_i)}{R_i^{n+1}}~, 
\end{equation} and 
the $Z_{n\alpha}$'s are the spherical
harmonics containing $\sin\phi$ or $\cos\phi$~\cite{Hutchings}. To get the crystal
field Hamiltonian $V_c$, one must sum this energy over all of the valence
electrons of the holmium (Ho$^{3+}$) moments, hence we have:
\begin{equation} 
V_{C} = -e\sum_j V(\mathbf{r}_j)~. 
\end{equation}
According to arguments provided by Stevens~\cite{Stevens} for evaluating the matrix elements 
of the crystal field Hamiltonian between wave functions specified by the angular momentum 
$\mathbf{J}$, the crystal field Hamiltonian 
 can be written in term of Stevens' operator equivalents $O_n^\alpha$,
 built out of the vector components 
 of $\mathbf{J}$ operators,
 \begin{equation} 
 V_{C}=\sum_{n,\alpha}B_n^{\alpha}O_n^{\alpha}~.
 \label{genCFH}
 \end{equation}
The Stevens' equivalent operators act on the angular momentum states 
of the wave functions. The matrix element of the radial part 
of the wave function is incorporated in the $B_n^\alpha$ parameters,
usually determined by fitting to  experimental (e.g. spectroscopic) data~\cite{Ronnow,Malkin,Hansen}.
From  angular momentum algebra, in the case of $4f$ electrons, we need to consider only $n=0,2,4,6$
 in the sum (\ref{genCFH}).
 
The choice of $B_{n}^{\alpha}$ coefficients in Hamiltonian (\ref{genCFH}) that 
do not vanish and have nonzero corresponding matrix elements 
is dictated by the point symmetry group of the
crystalline environment. The details
of the method and conventions for expressing the crystal field Hamiltonian
can be found in the review paper by Hutchings \cite{Hutchings}.
The point group symmetry of LiHoF$_4$ has 
 $S_4$ symmetry,
which means the lattice is invariant respect to a $\frac{\pi}{2}$ rotation about the $z$ axis 
and reflection with respect to the $x-y$ plane. The crystal field Hamiltonian for 
LiHoF$_4$ is therefore written as
\begin{eqnarray}\label{CrystalField} 
V_{C}&=& B_2^0 O_2^0 + B_4^0 O_4^0 + B_4^{4C} O_4^{4C}
 + B_4^{4S}O_4^{4S}\nonumber\\ &&+
B_6^0 O_6^0 + B_6^{4C} O_6^{4C} + B_6^{4S} O_6^{4S}. 
\end{eqnarray}
The
relevant  $B_{n}^{\alpha}$ crystal field parameters must be determined experimentally.
The relevant
operator equivalents are given in terms of angular momentum
operators~\cite{Hutchings} ($J_z$, $J_+$, $J_-$, $J^2$) by
\begin{eqnarray}
 O_2^0~~&=& 3J_z^2 - J^2 ~,\nonumber\\
 O_4^0~~&=& 35Jz^4 -30J^2J_z^2 + 25J_z^2 -6J^2 + 3J^4 ~,\nonumber\\
 O_4^{4C}&=&\frac{1}{2}(J_+^4 + J_-^4)~, \nonumber\\
 O_4^{4S}&=&\frac{i}{2}(J_+^4 - J_-^4) ~,\nonumber\\
 O_6^0~~&=& 231J_z^6 - 315J^2J_z^4 + 735J_z^4+ 105J^4J_z^2\nonumber\\
 &&  -
525J^2J_z^2 + 294J_z^2 - 5J^6 + 40J^4 -60J^2 ~,\nonumber\\
 O_6^{4C} &=& \frac{1}{4}(J_+^4 + J_-^4)(11J_z^2 - J^2 -38) + \mathrm{H.c.}~, {\rm and} \nonumber\\
 O_6^{4S} &=& \frac{1}{4i}(J_+^4 -J_-^4)(11J_z^2 - J^2 - 38) +
\mathrm{H.c.} 
\end{eqnarray}

The $B_n^\alpha$ parameters are chosen such that the resulting energy levels match those determined from 
spectroscopic data.
Two different set of experimentally determined crystal field parameters are given 
in Table~\ref{CrystalFieldParameters}. The first set of the parameters was determined by 
inelastic neutron scattering 
reported in Ref.~[\onlinecite{Ronnow}] and implemented in the calculations of Ref.~[\onlinecite{prabuddha}].
The next set of $B_n^\alpha$ parameters were determined using electron paramagnetic resonance
 (EPR) spectroscopy, and reported in a recent work~\cite{Malkin}.
 \begin{table}
\begin{center}
\begin{tabular}{crr}
\\
\\
  \hline
  \hline
  Parameter & Ref.~[\onlinecite{Ronnow}] & Ref.~[\onlinecite{Malkin}]  \\
  \hline
 $B_2^0$    & $-0.696$ K             & $ -0.609$ K\\
  $B_4^0$    & $4.06\times 10^{-3}$ K & $3.75\times 10^{-3}$ K\\
  $B_4^{4C}$ & $4.18\times10^{-2}$ K  & $ 3.15\times10^{-2}$ K\\
  $B_4^{4S}$ & 0 K                    & $ 2.72\times10^{-2}$ K\\
  $B_6^0$    & $4.64\times 10^{-6}$ K & $6.05\times10^{-6}$ K\\
  $B_6^{4C}$ & $8.12\times 10^{-4}$ K & $6.78\times10^{-4}$ K\\
  $B_6^{4S}$ & $1.137\times 10^{-4}$ K& $4.14\times10^{-4}$ K\\
  \hline
  \hline
\end{tabular}
\end{center}
\caption[]{The first column is the crystal field parameters (CFP) for
LiHoF$_4$ determined experimentally by fitting the results of random phase approximation 
spin-wave
dynamics calculation to neutron scattering data from
Ref.~[\onlinecite{Ronnow}]. The second column is the crystal field
parameters estimated using electron paramagnetic resonance (EPR) spectroscopy experiment~\cite{Malkin}. 
}
\label{CrystalFieldParameters}
\end{table}
 \section{}\label{B}
 In this Appendix, starting from Eq.~(\ref{cumulant}), we give the details of the derivation of 
 the effective perturbative Hamiltonian $H_{\rm eff}\left[ \psi_i\right]$ by cumulant expansion, when quantum fluctuations are small. 
 Deriving $H_{\rm eff}\left[ \psi_i\right]$, as defined by Eq.~(\ref{classical_part}) 
 one can rewrite the partition function of
 the system in a classical form.
 
 Referring to Eq.~(\ref{cumulant}), recalling that $|\psi\rangle$, is a direct product of $\sigma_i^z$ eigenstates, 
 the expectation value 
$\langle \psi\vert \sigma_x\vert\psi\rangle$ is zero, 
so $\langle\psi|\mathcal{H}|\psi\rangle=\langle\psi|H_0|\psi\rangle$.
Defining $E_0(\psi)\equiv\langle\psi\vert H_0\vert\psi\rangle$, we can write
$\langle\psi\vert\left(\mathcal{H}-\langle\psi|\mathcal{H}|\psi\rangle\right) ^n\vert\psi\rangle=\langle\psi\vert( \mathcal{H}-E_0(\psi))^n\vert\psi\rangle$.\\
Performing a polynomial expansion on $\left(\mathcal{H}-E_0(\psi)\right)^n
=\left[\left(H_0-E_0(\psi)\right)+H_1\right] ^n$, and
keeping terms to order of $O(\Gamma^2)$ in the polynomial expansion~($H_1\propto\Gamma$), we have
\begin{widetext}
\begin{eqnarray}
\langle\psi\vert\left(\mathcal{H}-E_0(\psi)\right) ^n\vert\psi\rangle &=&
\langle\psi\vert\left[\left(H_0-E_0(\psi)\right)+H_1\right] ^n\vert\psi\rangle\nonumber\\
 & = & \sum_{n_1,n_2,n_3}\delta(n_1+n_2+n_3,n-2)\times
\left[\langle\psi\vert(H_0-E_0(\psi))^{n_1}H_1(H_0-E_0(\psi))^{n_2}H_1
(H_0-E_0(\psi))^{n_3}\vert\psi\rangle\right]\nonumber\\
& = & \langle\psi\vert H_1\left[ H_0-E_0(\psi)\right]^{n-2}H_1\vert\psi\rangle
~.
\label{orderGamma}
\end{eqnarray}
\end{widetext} 

To write Eq.~(\ref{orderGamma}) we have used the fact that 
\begin{eqnarray}
\langle\psi|\left(H_0-E_0(\psi)\right) ^n|\psi\rangle=0
\end{eqnarray}
and
\begin{eqnarray}
\langle\psi\vert(H_0-E_0(\psi))^m H_1(H_0-E_0(\psi))^k\vert\psi\rangle=0~,
\end{eqnarray}
 for integer numbers $m$ and $k$. 
The effect of $\sigma_i^x$ on $\vert\psi\rangle$ is to flip the spin $i$. We define 
$\sigma_i^x\vert\psi\rangle=\vert f_i\psi\rangle$, where $f_i\psi$ means 
that the $i$'th spin has 
flipped, such that if the i'th spin was in the $| \! \uparrow\rangle$ 
or the $| \!\downarrow\rangle$ eigenstate
of $\sigma_i^z$, it changes into the $| \! \downarrow\rangle$ or  $| \! \uparrow\rangle$ state respectively.
 In Eq.~(\ref{orderGamma}), using $H_1=-\Gamma\sum_i\sigma_i^x$, we get
 \begin{eqnarray}
  &\langle\psi\vert H_1\left[ H_0-E_0(\psi)\right] ^{n-2}H_1\vert\psi\rangle\nonumber\\
  \hspace{-15mm}&=\Gamma^2\sum_{i,j}\langle\psi\vert\sigma_i^x\left[ H_0-E_0(\psi)\right] ^{n-2}
  \sigma_j^x\vert\psi\rangle\nonumber\\
  &=\Gamma^2\sum_{i,j}\langle f_i\psi\vert\left[ H_0-E_0(\psi)\right] ^{n-2}
  \vert f_j\psi\rangle~.
  \label{flipflip}
 \end{eqnarray}
 Here $\langle f_i\psi\vert\left[ H_0-E_0(\psi)\right] ^{n-2}
  \vert f_j\psi\rangle$ is zero, unless $i=j$. 
Thus, Eq.~(\ref{orderGamma}) can be written as
\begin{eqnarray}
\langle\psi\vert\left( \mathcal{H}-E_0(\psi)\right) ^n\vert\psi\rangle=
\Gamma^2\sum_i\left[E_0(f_i\psi)-E_0(\psi)\right]^{n-2}.~~
\end{eqnarray} 
Considering the definition of $H_{\rm eff}$, by substituting $E_0(f_i\psi)-E_0(\psi)=2(h_i+h_0)\sigma_i^z$ in  
Eq.~(\ref{cumulant}), 
we obtain
\begin{eqnarray}
H_{\rm eff}&=&
H_0-\beta\Gamma^2\sum_i\sum_{n>1}^{\infty}\frac{1}{n!}\left[ -2\beta(h_i+h_0)\right]^{n-2}\nonumber\\
&=& H_0+\beta\Gamma^2\sum_{i}\{\sigma_i^zF_1\left[ 2\beta(h_i+h_0)\right]\nonumber\\
&&- F_0\left[ 2\beta(h_i+h_0)\right]\}.
\label{eff_app} 
\end{eqnarray} 
  In Eq.~(\ref{eff_app}), $h_i$ is the total local field affecting the spin at site $i$
  by other spins , which is
\begin{eqnarray}
h_i=-\sum_{j\neq i}\mathcal{L}_{ij}^{zz}\sigma_j^z-\mathcal{J}_{\rm ex}\sum_{\rm NN}\sigma_{\rm \rm NN}^z~,
\label{localfieldb}
\end{eqnarray}
and $h_0$ is the external longitudinal field in the $z$ direction. 
The functions $F_0(x)$ and $F_1(x)$ are defined as
\begin{eqnarray}
F_0(x)=\frac{\cosh(x)-1}{x^2},\nonumber \\
F_1(x)=\frac{\sinh(x)-x}{x^2}.
\end{eqnarray}
\section{}\label{C}
In this Appendix, we establish the relationship between the real thermodynamical quantities as physical observables and their corresponding pseudo-operators, which are obtained using the perturbative
effective classical Hamiltonian of Eq.~(\ref{eff}). These thermodynamical quantities are calculated by employing
the derived pseudo-operators in our perturbative classical Monte Carlo simulations.

Writing the partition function in terms of the perturbative effective Hamiltonian
$H_{\rm eff}\left[ \psi_i\right]$, the pseudo-operators corresponding
to $\langle E \rangle$, $\langle M_z \rangle$, $\langle M_x \rangle$,  $\langle M_z^2 \rangle$, and
 $\langle M_z^4 \rangle$, which should be calculated to obtain thermodynamical quantities
using Monte Carlo simulations are written as
\begin{eqnarray}
\langle E \rangle  
&=& -\frac{1}{Z}\frac{\partial Z}{\partial \beta} =
\left\langle H_{\rm eff} + \beta \frac{\partial H_{\rm eff}}{\partial \beta}\right\rangle \\
 \langle M_z \rangle &=& \left\langle- \frac{\partial H_{\rm eff}}{\partial h_0}\right\rangle\\
 \langle M_x \rangle &=& \left\langle- \frac{\partial H_{\rm eff}}{\partial \Gamma}\right\rangle\\
 \langle M_z^2 \rangle &=& \left\langle \left( \frac{\partial H_{\rm eff}}{\partial h_0}\right)^2 
  -\frac{1}{\beta}\frac{\partial^2 H_{\rm eff}}{\partial h^{2}_0}\right\rangle\label{Mz2} \\
 \langle M_z^4 \rangle 
&=&
\frac{1}{\beta^4}
\left\langle -\beta\frac{\partial^4 H_{\rm eff}}{\partial h_0^{4}} \right .
 +4\beta^2 \frac{\partial^3 H_{\rm eff}}{\partial h_0^{3}} \frac{\partial H_{\rm eff}}{\partial h_0} \\
  & & -6\beta^3\frac{\partial^2 H_{\rm eff}}{\partial h_0^{2}}
\left(\frac{\partial H_{\rm eff}}{\partial h_{0}} \right)^2
+3\beta^2\left(\frac{\partial^2 H_{\rm eff}}{\partial h_0^{2}}\right)^2 \nonumber \\
    & & +  \left . \beta^4\left(\frac{\partial H_{\rm eff}}{\partial h_0}\right)^4 \right\rangle ,
\end{eqnarray} 
 where $E$, $M_z$ and $M_x$ are the energy and magnetization in the $z$ and $x$ 
direction and their equivalent
 pseudo-operators which should be calculated are on right. The 
thermal average is denoted by $\langle\dots\rangle$. 
Applying the derivatives and using the perturbative effective Hamiltonian (\ref{eff}), 
 we find:
\begin{eqnarray}
E & = & E_0+2\beta\Gamma^2\sum_{i}\lbrace \sigma_i^zF_1\left[ 2\beta(h_i+h_0)\right]\nonumber\\
& & -F_0\left[ 2\beta(h_i+h_0)\right]\rbrace
\nonumber\\
& & + \beta\Gamma^2\sum_{i}2\beta(h_i+h_0)\lbrace \sigma_i^z F^{(1)}_1\left[ 2\beta(h_i+h_0)\right]\nonumber\\
& & - F^{(1)}_0\left[ 2\beta(h_i+h_0)\right]\rbrace~, 
\end{eqnarray} 
while $M_x$ is
\begin{eqnarray}
M_x & = & -2\beta\Gamma\sum_{i}\lbrace \sigma_i^zF_1\left[ 2\beta(h_i+h_0)\right]\nonumber\\ & &- F_0\left[ 2\beta(h_i+h_0)\right]\rbrace
\end{eqnarray} 
and $M_z$
\begin{eqnarray}
M_z & = &\label{Mz4} \sum_i\sigma_i^z-2\beta^2\Gamma^2\sum_{i}\lbrace \sigma_i^zF^{(1)}_1\left[ 2\beta(h_i+h_0)\right]\nonumber\\ & & - F^{(1)}_0\left[ 2\beta(h_i+h_0)\right]\rbrace~,
\end{eqnarray} 
with $F^{(n)}_i(x)$  defined as $F^{(n)}_i=\frac{d^nF_i(x)}{dx^n}$, where $i=1,0$.\\
In order to find an expression for $\langle M_z^2\rangle$ and $\langle M_z^4 \rangle$, we need to calculate 
$\frac{\partial H_{\rm eff}}{\partial h_0}$, $\frac{\partial^2 H_{\rm eff}}{\partial h_0^2}$, 
$\frac{\partial^3 H_{\rm eff}}{\partial h_0^3}$, and $\frac{\partial^4 H_{\rm eff}}{\partial h_0^4}$~.
We find:
\begin{eqnarray}
\frac{\partial H_{\rm eff}}{\partial h_0} & = & -\sum_i\sigma_i^z+2\beta^2\Gamma^2\sum_{i}\lbrace \sigma_i^z F^{(1)}_1\left[ 2\beta(h_i+h_0)\right]\nonumber\\ & & 
- F^{(1)}_0\left[ 2\beta(h_i+h_0)\right]\rbrace~,
\end{eqnarray}
 \begin{eqnarray}
 \frac{\partial^2 H_{\rm eff}}{\partial h_0^2} & = & 
 4\beta^3\Gamma^2\sum_{i}\lbrace \sigma_i^z F^{(2)}_1\left[ 2\beta(h_i+h_0)\right]\nonumber\\ 
 & & - F^{(1)}_0\left[ 2\beta(h_i+h_0)\right]\rbrace~,
 \end{eqnarray}
 \begin{eqnarray}
 \frac{\partial^3 H_{\rm eff}}{\partial h_0^3} & = & 8\beta^4\Gamma^2\sum_{i}\lbrace \sigma_i^z F^{(3)}_1\left[ 2\beta(h_i+h_0)\right]\nonumber\\ 
 & & - F^{(3)}_0\left[ 2\beta(h_i+h_0)\right]\rbrace~~~\text{and}
 \end{eqnarray}
 \begin{eqnarray}
 \frac{\partial^4 H_{\rm eff}}{\partial h_0^4} & = & 16\beta^5\Gamma^2\sum_{i}\lbrace \sigma_i^z F^{(4)}_1\left[ 2\beta(h_i+h_0)\right]\nonumber\\
 & &- F^{(4)}_0\left[ 2\beta(h_i+h_0)\right]\rbrace~.
 \end{eqnarray}



\begin{thebibliography}{99}
\bibitem{Sachdev}
S. Sachdev, {\it  Quantum Phase Transitions},  (Cambridge University Press,1999).
\bibitem{Sondhi}
S. L. Sondhi, S. M. Girvin, J. P.  Carini, and D. Shahar, Rev. Mod. Phys. {\bf 69}, 315 (1997).
\bibitem{Pfeuty} 
R. J. Elliott, P. Pfeuty, and C. Wood,
Phys. Rev. Lett. {\bf 25}, 443 (1970).
\bibitem{Sen}
B. K. Chakrabarti, A. Dutta, and P. Sen,
{\it
Quantum Ising Phases and Transitions in Transverse Ising Models},
(Springer-Verlag, Heidelberg, 1996).
\bibitem{deGennes}
P. G. de Gennes, Solid State Comm. \textbf{1}, 132 (1963).
\bibitem{Binder}
 K. Binder and A. P. Young, Rev. Mod. Phys. {\bf 58}, 801 (1986). 
\bibitem{Ballestros}
  H. G. Ballesteros {\it et al.}
 Phys. Rev. B {\bf 62}, 14237 (2000).
\bibitem{Rieger-PRL}
H. Rieger and A. P. Young, Phys. Rev. Lett. {\bf 72}, 4141 (1994).
 \bibitem{Rieger-PRB}
 H. Rieger and A. P. Young, Phys. Rev. B {\bf 54}, 3328 (1996).
\bibitem{Guo}
M. Guo, R. N. Bhatt, and D. A. Huse, Phys. Rev. B {\bf 54}, 3336 (1996).
\bibitem{Griffiths}
 R. B. Griffiths, Phys. Rev. Lett. {\bf 23}, 17 (1969).
\bibitem{McCoy}
B. M. McCoy,  Phys. Rev. Lett. {\bf 23}, 383 (1969).
\bibitem{Bitko}
 D. Bitko, T. F. Rosenbaum, and G. Aeppli, Phys. Rev. Lett. \textbf{77}, 940 (1996).
 \bibitem{Wu}
W. Wu, B. Ellman, T. Rosenbaum, G. Aeppli, and D. H. Reich. Phys. Rev. Lett. {\bf{67}}, 2076 (1991).
\bibitem{Ronnow} 
 H. M. R\o nnow, J. Jensen, R. Parthasarathy, G. Aeppli, T. F. Rosenbaum, D. F. McMorrow, and C. Kraemer,
 Phys. Rev. B \textbf{75}, 054426 (2007).
 \bibitem{Hansen} 
 P. E. Hansen, T. Johansson, and R. Nevald, Phys. Rev. B {\bf 12}, 5315 (1975).
\bibitem{Eastham}
A. Chin and P. R. Eastham, cond-mat/0610544.
 \bibitem{Tabei-Vernay}
 S. M. A. Tabei, F. Vernay, M. J. P. Gingras, arXiv:0708.2286 (to appear in Phys. Rev. B).  
 \bibitem{prabuddha}
P. B. Chakraborty, P. Henelius, H. Kj\o nsberg,
 A. W. Sandvik, and
  S. M. Girvin, Phys. Rev. B \textbf{70}, 144411 (2004).
 \bibitem{Silevitch}
 D. M. Silevitch, D. Bitko, J. Brooke, S. Ghosh, G. Aeppli, and T. F. Rosenbaum, Nature {\bf 448}, 567 (2007).
  \bibitem{Reich}
D. H. Reich, B. Ellman, J. Yang, T. F. Rosenbaum, G. Aeppli, and D. P. Belanger,
Phys. Rev. B {\bf 42}, 4631 (1990).
\bibitem{No-SG}
We note that the existence of a spin glass transition in LiHo$_x$Y$_{1-x}$F$_4$ (x=0.167)
 in \textit{zero} transverse magnetic field has very recently been questioned. See Ref.~[\onlinecite{Jonsson}]
\bibitem{Jonsson}
 P. E. J\"onsson, R. Mathieu, W. Wernsdorfer, A. M. Tkachuk, B. Barbara, Phys. Rev. Lett. {\bf 98}, 256403 (2007).
\bibitem{Brooke-thesis}
J. Brooke,
Ph.D. thesis, U. Chicago (2000).
\bibitem{Mydosh}
J. A. Mydosh, {\it Spin Glasses: An Experimental Introduction}, (Taylor \& Francis, London,1993).
\bibitem{Wu-thesis}
W. Wu, Ph.D. thesis, U. Chicago (1992).

\bibitem{Schechter-PRL2}
M. Schechter and N. Laflorencie, Phys. Rev. Lett. {\bf 97}, 137204 (2006).
\bibitem{Tabei-PRL}
S. M. A. Tabei, M. J. P. Gingras, Y.-J. Kao, P. Stasiak, and J.-Y. Fortin, 
Phys. Rev. Lett. {\bf 97},  237203 (2006).
\bibitem{Schechter-JPC}
 M. Schechter, P.C.E Stamp, and N. Laflorencie, J. Phys.: Condens. Matter {\bf 19}, 145218 (2007).
\bibitem{Ghosh-Science}
 S. Ghosh, R. Parthasarathy, T. F. Rosenbaum, and G. Aeppli, Science {\bf 97},
 2195 (2002). 
\bibitem{Ghosh-Nature}
 S. Ghosh, T. F. Rosenbaum, G. Aeppli, and S. N. Coppersmith, Nature, {\bf 425}, 48
 (2003). 
 \bibitem{Stephen}
 M. J. Stephen and A. Aharony, J. Phys. C {\bf 14}, 1665 (1981). 
 \bibitem{Snyder}
 J. Snider and C.C. Yu,
  Phys. Rev. B {\bf 72}, 214203 (2005).
\bibitem{Biltmo}
A. Biltmo and P. Henelius, Phys. Rev. B {\bf 76}, 054423 (2007).
 
\bibitem{sandvik}
  A. W. Sandvik and J. Kurkij\aa rvi, Phys. Rev. B \textbf{43}, 5950 (1991);
 A. W. Sandvik, Phys. Rev. E \textbf{68}, 056701 (2003).
 \bibitem{Schechter}
 M. Schechter and P. C. E. Stamp
 Phys. Rev. Lett. \textbf{95}, 267208 (2005).
 \bibitem{Malkin}
G. S. Shakurov, M. V. Vanyunin, B. Z. Malkin, B. Barbara, R. Yu. Abdulsabirov, and S. L. Korableva,
 Appl. Magn. Reson., \textbf{28}  251 (2005).
 
\bibitem{conditionally}
 L. N. Kantorovich and I. I. Tupitsyn, J. Phys.: Condens. Matter \textbf{11}, 6159 (1999). 
\bibitem{comment}
The sum of an infinite number of dipole-dipole interactions
 is conditionally convergent
and depends on the order of the summation. For example,
 if the dipole-dipole interactions of a central unit cell with unit cells located on an 
 ever-increasing long needle-shaped sample, the energy converges to a different value than if the interaction
  energies had been summed spherically. Roughly speaking, this conditional convergence arises
  because the number of interacting dipoles on a shell of radius $R$ grows like $R^2$,
  while the strength of a single dipole-dipole interaction falls like $1/R^3$, and the mathematical
  $\sum_{n=1}^{\infty}\frac{1}{n}$ summation diverges. 
 The value that the sum converges to, depends on the shape of the boundary
 of the system.
In the present work, the effect of the geometry of the boundary is incorporated in Ewald summation
 technique. See Ref.~[\onlinecite{conditionally}]
 \bibitem{Luttinger}
J. M. Luttinger and L. Tisza, Phys. Rev. \textbf{70}, 954 (1946). 
\bibitem{Griffiths2}
R. Griffiths, Phys. Rev. \textbf{176} (1968), 655. 
 \bibitem{RF}
 J. A. Barker and R. O. Watts, Mol. Phys. \textbf{26}, 789 (1973).
 \bibitem{Zeeman}
 J. M. Ziman, {\it Principles of the Theory of Solids} (Cambridge Univ. Press, Cambridge, ed. 2, 1972).
\bibitem{huang}
M. Born, S. Huang, {\it Dynamical Theory of Crystal Lattices} (Oxford Univ. Press, New York, 1968).
\bibitem{Leeuw}
S .W. de Leeuw, J. W. Perram, and E. R. Smith, Ann. Rev. Phys. Chem. \textbf{37}, 245 (1986).
\bibitem{spinice}
 For example, the Ewald summation method has proved quite efficient to allow a characterization of the 
 thermodynamic properties of rare-earth spin ice materials, such as Ho$_2$Ti$_2$O$_7$ and Dy$_2$Ti$_2$O$_7$, 
 and a determination of the exchange in these materials. 
 See Ref.~[\onlinecite{Melko}] and references therein.
\bibitem{Melko}
 R. G. Melko, and M. J. P. Gingras, J. Phys.: Condens. Matter {\bf 16}, R1277 (2004).
 \bibitem{JensenBook}
 J. Jensen and A. R. Mackintosh,
{\it Rare Earth Magnetism} (Oxford Univ. Press, Oxford, 1991).
\bibitem{SSE}
The quantum Monte Carlo method (QMC) based on stochastic series expansion (SSE) of Ref.~[\onlinecite{sandvik}]
amounts to a numerical summation of $\beta H$ to high powers, where $H$ is the Hamiltonian and $\beta=1/k_{\rm B}T$.
The method we use splits $H$ into the classical Ising sector, $H_0$, and the quantum transverse field term, $H_1$,
and resums ``analytically'' all the terms in $\beta H_0$ and retains only the leading $(\beta H_1)^2$ 
term in evaluating thermodynamic averages.   
 \bibitem{Mennenga}
 G. Mennenga, L. J. de Jongh, and W. J. Huiskamp, J. Magn. Magn. Matter. {\bf{44}}, 59 (1984).
 \bibitem{Beauv}
 P. Beauvillain, J. P. Renard, I. Laursen, and P. J. Walker, Phys. Rev. {\bf{B}} 18, 3360 (1978).
 \bibitem{11K} 
In LiHoF$_4$ the value of the energy gap $\Delta$ between the ground state doublet and the excited state for $B_x=0$, 
strongly depends on the crystal field Hamiltonian. Since there is an ambiguity in the crystal field parameters
among different experimental works, there is also an ambiguity in the calculated energy gap. For different estimations 
of $\Delta$ see Refs.~[\onlinecite{Hansen,Ronnow, Malkin}]
 \bibitem{Magarino}
 J. Magarino, J. Tuchendler, P. Beauvillain, and I. Laursen, Phys. Rev. B \textbf{21}, 18 (1980).
 \bibitem{LiTb}
 L. M. Holmes, J. Als-Nielsen, and H. J. Guggenheim, Phys Rev. B \textbf{12}, 180 (1974).
 
 \bibitem{precumulant}
 We follow closely the method laid out in Ref.~[\onlinecite{cumulant}] as well as
 adopt their notation. However, we provide somewhat more details to assist the reader.
  \bibitem{cumulant}
R. J. Creswick, H. A. Farach, J. M. Knight, and C.P. Poole 
Phys Rev. B \textbf{38}, 4712 (1988).
\bibitem{cum}
M. Le Bellac, {\it Quantum and Statistical Field Theory} (Oxford Univ. Press, New York, 1992).
\bibitem{trotter} 
M. Suzuki, Prog. Theor. Phys. {\bf{46}} 1337 (1971); \textit{Quantum Monte 
Carlo Methods}, Ed. M. Suzuki Springer-Verlag, Heidelberg (1987).
\bibitem{rieger}
H. Rieger and N. Kawashima, Eur. Phys. J. B {\bf 9} 233 (1999).
\bibitem{cooke}
 A. H. Cooke, D. A. Jones, J. F. A. Silva, and M. R. Wells,
J. Phys. C: Sol. St. Phys. \textbf{8}, 4083 (1975).
\bibitem{Battison} 
J. E. Battison, A. Kasten, M. J. M. Leask, J. B. Lowry,
and B. M. Wanklyn, J. Phys. C: Sol. St. Phys. \textbf{8}, 4089
(1975).
\bibitem{Jackson}
 J. D. Jackson, {\it Classical Electrodynamics} (Wiley, New York, 1998), 3rd ed.
 \bibitem{Xu}
H. J. Xu, B. Bergersen, and Z. Racz,  J. Phys.:
 Condens. Matter \textbf{4},  2035 (1992).
\bibitem{vacuum_comment}
 The magnetic permeability $\mu'$ is usually not known beforehand. In general as discussed 
in Ref~[\onlinecite{Leeuw}] it can be estimated rigorously in a self consistent way.
In the problem  that we are interested in here, we simulate an isolated sphere,
 which the effect of the infinite continuum surrounding it is 
incorporated via $B_z^{\rm sph}$ defined in Eq.~(\ref{demagnet}). 
We are interested in the situation where 
$\chi_{\rm sph}=3/4\pi$.  When this situation is fullfilled the whole long needle-shaped bulk
 is in the paramagnetic regime where in the thermodynamic limit the macroscopic magnetization 
 of the bulk is zero. Therefore, it seems quite reasonable
to consider $\mu'$=1 for the spherical sample for which the simulation
is being carried
without embarking into
complex self-consistent calculations, which are beyond 
the scope of this paper, specially since at the end
the results so obtained are 
consistent with those
found when considering a long needle-shaped 
bulk with no demagnetization effects.
 
 \bibitem{Tupizin}
 I. I. Tupizin and I. V. Abarenkov Phys. Status Solidi B \textbf{82}, 99 (1977).
 \bibitem{binder}
 K. Binder, Z. Phys. B \textbf{43}, 119 (1981).
\bibitem{Larkin}
A. I. Larkin and D.E. Khmel'nitskii, Soviet Physics JETP \textbf{29}, 1123 (1969).
 \bibitem{Pawel}
 P. Stasiak and M. J. P. Gingras, \textit{unpublished}.
 \bibitem{Hutchings}
 M. T. Hutchings, Solid State Phys. {\bf 16}, 227 (1964).
\bibitem{Stevens}
 K. W. H. Stevens, Proc. Phys. Soc. A {\bf 65}, 209 (1952).
 
 
%
\end{thebibliography}
\end{document}